\documentclass{article}

\pdfoutput=1

\usepackage{amsmath,amssymb,amsfonts,textcomp}
\usepackage{physics}
\usepackage{graphicx}
\usepackage{wrapfig}
\usepackage{float,flafter}
\usepackage{hyperref}
\usepackage{inputenc}
\usepackage{caption}
\usepackage{subcaption}
\usepackage{tcolorbox}
\usepackage[a4paper,left=30mm,top=30mm,right=25mm,bottom=30mm]{geometry}

\usepackage[
backend=biber,
style=numeric,
sorting=none,
maxbibnames=5
]{biblatex}
\newcommand{\wnh}{\omega_\text{NH}}
\newcommand{\wh}{\omega_\text{H}}
\newcommand{\nas}{\text{NA}_S}
\newcommand{\feq}{f_{eq}}
\newcommand{\disc}{\text{Disc}}
\newcommand{\N}{\mathcal{D}}
\newcommand{\teff}{\tau_\text{eff}}

\title{New insights into the analytic structure of correlation functions via kinetic theory}
\author{Robbe Brants$^1,$\footnote{\href{mailto:robbe.brants@ugent.be}{robbe.brants@ugent.be}}\vspace{5pt} \\
        $^1$\em{\small Department of Physics and Astronomy, Ghent University, 9000 Ghent, Belgium}}
\date{}

\begin{document}

\maketitle

\begin{abstract}
The way a relativistic system approaches fluid dynamical behaviour can be understood physically through the signals that will contribute to its linear response to perturbations. What these signals are is captured in the analytic structure of the retarded correlation function. The non-analyticities can be grouped into three types based on their dimension in the complex frequency plane. In this paper, we will use kinetic theory in the (momentum dependent) relaxation time approximation to find how we can calculate their corresponding signals. In the most general case of a system with particles that have a continuum of thermalization rates, we find that a non-analytic region appears. To calculate its signal, we introduce the {\em non-analytic area density} that describes the properties of this region, and we construct a method to calculate it.  Further, to take into account the ambiguity present in signal analysis, following from manipulations of the non-analyticities, we will identify two specific choices called {\em pictures} with interesting analytic properties and compare in what scenarios each picture is most useful. 
\end{abstract}

\section{Introduction}
In the last two decades, ultrarelativistic heavy ion collisions at the Relativistic Heavy Ion Collider (RHIC) and Large Hadron Collider (LHC) have been used  to create and study the quark-gluon plasma, an equilibrium phase of QCD \cite{STAR:2005gfr, ALICE:2008ngc}. The evolution from these far-from-equilibrium collisions towards this relativistic fluid has constituted a powerful motivation for theoretical studies of equilibration processes involving relativistic particles \cite{Busza:2018rrf}. As a result, research on thermalization towards relativistic fluids has progressed tremendously. This research is typically based on ab initio simulations of real time evolution where it is difficult to gain physical intuition from the results \cite{Berges:2020fwq}. To further advance the field, it is therefore useful to approach the problem from a more fundamental point of view, where we investigate regimes in which the equations that govern time evolution are heavily simplified, and use the results to interpret the full ab initio simulations. \\
\\
One example of such a regime is the linear regime, where we study the retarded correlation function $G(t, \vec{r})$. This function gives the linear response of a conserved quantity to a small perturbation. To find the dominant plane waves, we solve the equations of motion in Fourier space and calculate the correlation function $G(\omega,\vec{k})$. For exponentially decaying waves, the values of $\omega$ will be complex numbers. The signals that will contribute to the real-time correlation function are then given by singularities in the analytic continuation of $G(\omega,\vec{k})$ towards complex $\omega$. The drawback of looking at complex $\omega$ is that there is now a certain ambiguity, as multiple analytic continuations can result in the same total signal in real time. Each of the non-analyticities has a characteristic signal determined by its location and the distribution of its non-analyticity. Afterwards, the signals are compared to the original simulations and provide a physical interpretation. Signal analysis can also be used to directly compare different theories, since these signals are purely physical properties and their interpretation does not depend on the underlying microscopic theory used to calculate them. \\
\\
The theory we focus on to find these signals is kinetic theory, because it allows all different types of non-analyticities to be present and can be used to describe both the hydrodynamic and non-hydrodynamic sector. We will therefore be able to predict properties of the system for a wide range of timescales. It is also capable of modelling QCD dynamics at weak coupling, allowing a connection to the experiments previously mentioned. In kinetic theory, a system containing on-shell particles is described entirely through a single statistical distribution function $f(t, \vec{r}, \vec{p})$. The evolution of the distribution function is determined by the Boltzmann equation. In the Relaxation Time Approximation (RTA) \cite{Anderson:1974nyl}, the collision kernel $C[f]$ of the relativistic Boltzmann equation is linearized around an equilibrium distribution $\feq$. Although a straightforward model, RTA can be used to understand the transition from an out-of-equilibrium system towards the hydrodynamic regime \cite{Baym:1984np, Florkowski:2013lya}. The advantage of this simplified theory is that it can be solved analytically in Fourier space, leading to an exact signal analysis. It was therefore also the first model that was used to understand the analytic structure of correlation functions in kinetic theory. This was done for the case where $\tau_R$ is a constant, leading to a horizontal branch cut in the analytic structure \cite{Romatschke_2016}. The analysis was then extended to a momentum dependent relaxation time \cite{Kurkela:2017xis}, where the analytic structure was deformed into 2 vertical branch cuts. To ensure energy-momentum conservation, the temperature and macroscopic velocity had to be redefined. A more physical solution is given by adapting the collision kernel while keeping the matching conditions for the thermodynamic variables general \cite{Rocha:2021zcw}. \\
\\
However, there is one type of signal that has barely been addressed so far. When moving from standard to momentum dependent RTA, the horizontal branch cut gets spread out into an entire 2-dimensional region with non-analytic properties. Previously, this region was avoided by manipulating the non-analyticity into a picture with 2 vertical branch cuts instead \cite{Kurkela:2017xis}. There is a lot of physics captured in the non-analytic region however, making it very much worth investigating. Furthermore, in numerical calculations the momentum grid is typically real and no explicit manipulations are possible, so this non-analytic region is mostly unavoidable for general kernels. Such a numerical method is described in \cite{Ochsenfeld:2023wxz} and the results show how the non-analytic region is formed for a theory with dissipation and without a fixed relaxation time. This is a physical scenario, since a general quantum field theory has gapless non-analyticities, as proven in \cite{gavassino2024gapless}. \\
\\
These recent results motivate us to analytically calculate the analytic structure of momentum dependent RTA, with a relaxation time that can grow in an unrestricted way as the momentum of the particles increases. Since the signals of the non-analyticities depend on what quantities are conserved and how they are perturbed, we have to choose a channel we want to investigate. This channel specifies what thermodynamic variables are compared in the retarded correlation function and how they were perturbed. Because most results are analogous for all channels, we will be working with the simplest and most intuitive physical channel: particle (number) diffusion. Here the total number of particles is kept constant and we look at how the particle density evolves. \\
\\
The general goal of this paper is to find descriptions of the different types of non-analyticities in full detail and interpret them physically. We expect that a full signal analysis is possible purely based on the distribution of the non-analyticity. To do this, we first have to construct a way of calculating the signals analytically. Since there are multiple equivalent analytic structures, we identify 2 distinct pictures and compare them. We also want to investigate whether the non-analytic distribution can be calculated entirely through numerical methods.\\
\\
To achieve these goals, the paper consists of the following parts: in section \ref{sec:physics behind nonan} we will introduce the possible types of non-analyticities, how they can occur in the analytic structure, and what their physical interpretation is. Knowing what they are, we first start by fully describing the non-analyticities in standard RTA in section \ref{sec:standard} and constructing the pictures that will be relevant. Finally we complete our analysis with the more general momentum dependent RTA in \ref{sec:pdep}, where we look at the behaviour of the modes and calculate the non-analytic density. Appendix \ref{ap: rules} contains mathematical rules that are required for calculating the non-analyticity of a product of non-analytic functions. \\
\\
Throughout the paper we work in natural units $c=\hbar=k_B=1$. The convention for the Fourier transform in 3 spatial dimensions is $f(\omega,\vec{k}) = \int dtd^3\vec{r}e^{i\omega t-i\vec{k}\cdot\vec{r}}f(t,\vec{r})$. We will often use the notation $\omega = x-iy$ with $x$ and $y$ real numbers to describe non-analytic functions in frequency space. 

\section{The physics behind non-analyticities} \label{sec:physics behind nonan}
\begin{wrapfigure}{R}{0.38\textwidth}
\vspace{-10pt}
	\includegraphics[width=0.38\textwidth,keepaspectratio]{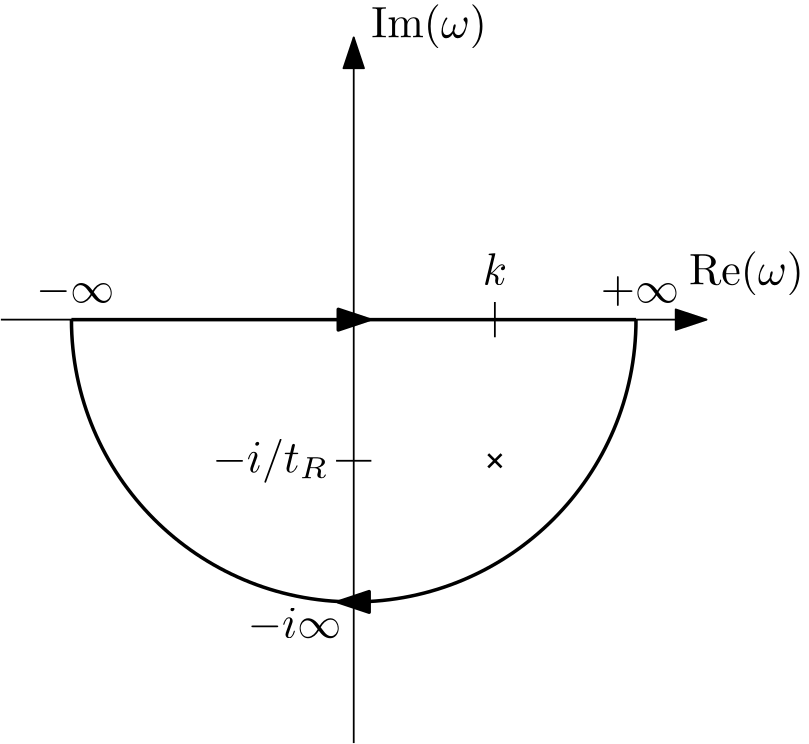}
	\caption{Contour integral for a function with pole at $k-i/t_R$.}
	\label{fig:pole}
 \vspace{-20pt}
\end{wrapfigure}

Since we will be working with a linear kernel, it is usually possible to find an exact solution in Fourier space. If we want to find the behaviour of the system in real time however, we have to apply an inverse Fourier transform. Integrating $\omega$ over the real axis is analytically often difficult and gives very little physical intuition. It is therefore recommended to connect the ending and starting point of the Fourier integral using a half-arc passing through negative imaginary infinity, creating a closed contour integral. This is shown with a diagram in \ref{fig:pole}. The contribution from this arc vanishes due to the factor $e^{-i\omega t}$, so the final result in real-time is equal to the sum over all non-analyticities in the lower complex plane. \\
\\
In this section we will go over all types of non-analyticities and their signals that can occur in kinetic theories and explain their physical interpretation. The goal is to build up intuition through a simplistic example in order to understand the correlation functions we will encounter in a more realistic and physical theory.

\subsection{Poles}

The simplest case one can consider is a zero-dimensional non-analyticity. This corresponds to a pole in the correlation function. For example, consider the following one-dimensional Boltzmann equation for the distribution function $f(t,z)$ with kernel $C[f]$, where the distribution function decays exponentially to zero with relaxation time $t_R$:
\begin{equation}
    \frac{\partial}{\partial t} f(t,z)+\frac{\partial}{\partial z} f(t,z) = C[f] = -\frac{1}{t_R}f(t,z)\,.
    \label{eq:nhBoltzmann}
\end{equation}
To find the analytic structure of $f(\omega)$ as a function of the initial state we apply a Laplace transform in time and a Fourier transform in the spatial coordinate:
\begin{equation}
    -i\omega f(\omega,k) - f(t=0,k) +ik f(\omega,k)= -\frac{1}{t_R}f(\omega,k)\,.
\end{equation}
For an initial perturbation at the origin $f(t=0,z)=f(0)\delta(z)$ we find $f(t=0,k)=f(0)$. This equation has the~solution
\begin{equation}
    f(\omega,k)= \frac{f(0) t_R}{1-i\omega t_R+ik t_R}\,.
    \label{eq:fwk}
\end{equation}
As a result we end up with a pole at $\omega_*=-i/t_R+k$ with residue $if(0)$. The inverse Laplace transform thus results in
\begin{equation}
    f(t,k)=\frac{1}{2\pi}\int_{-\infty}^\infty f(\omega) e^{-i\omega t} d\omega = -i\text{Res}(f, \omega_*) e^{-i\omega_* t} = f(0)e^{-t/t_R-ikt}\,,
\end{equation}
or in real-space $f(t,z)=f(0)e^{-t/t_R}\delta(z-t)$. We can conclude that a pole $\omega_*$ in the analytic structure corresponds to a mode whose dissipative properties are determined by Im$(\omega_*)$ while the real part adds propagation. Since the imaginary part of the pole is always $-1/t_R$, independent of the value of $k$, the mode is purely dissipative. 

Modes are the most basic signals that can be used to describe a system and form an indispensable part of the analytic structure in both weakly and strongly coupled theories \cite{Romatschke_2016,Hiscock:1987zz,Hartnoll:2005ju,Kovtun:2005ev}.

\subsection{Branch cuts}

In the previous section we did not take any momentum dependence into account and only had particles moving in the positive $z$ direction. For a system with 3 spatial dimensions however, the direction in which the distribution function $f(\vec{r},\vec{p},t)$ changes spatially depends on the momentum distribution. We can take this dependence by replacing the spatial derivative in (\ref{eq:nhBoltzmann}) by $\vec{v}\cdot\vec{\nabla}$ where $\vec{v}=\vec{p}/p$ is the direction the particle is moving in, and $p=|\vec{p}|$. As a result (\ref{eq:fwk}) gets modified to 
\begin{equation}
    f(\omega,k,\vec{p})= \frac{f(t=0) t_R}{1-i\omega t_R+i\vec{v}\cdot\vec{k} t_R}\,,
\end{equation}
and the pole now lies at $\omega_*=-i/t_R+\vec{v}\cdot\vec{k}$. We assume the initial perturbation $f(0,k,p)$ to be isotropic. Since $\vec{v}\cdot\vec{k}$ lies anywhere between and $k=|\vec{k}|$ and -$k$, for $k>0$ we do not have a single pole, but a line of non-analyticities at Im$(\omega)=-i/t_R$. To see how this results in a branch cut, we can integrate $f(\omega,k,\vec{p})$ over the angular part of $\vec{p}$:
\begin{equation}
    f(\omega,k,p) = \int_{-1}^{1} \frac{d(\cos\theta)}{2} \frac{f(0,k,p) t_R}{1-i\omega t_R+ik t_R\cos\theta} = \frac{if(0,k,p)}{2k} \log(\frac{i+\omega t_R+k t_R}{i+\omega t_R-k t_R})\,.
    \label{eq:f branch}
\end{equation}

\begin{wrapfigure}{R}{0.35\textwidth}
\vspace{-10pt}
	\includegraphics[width=0.35\textwidth,keepaspectratio]{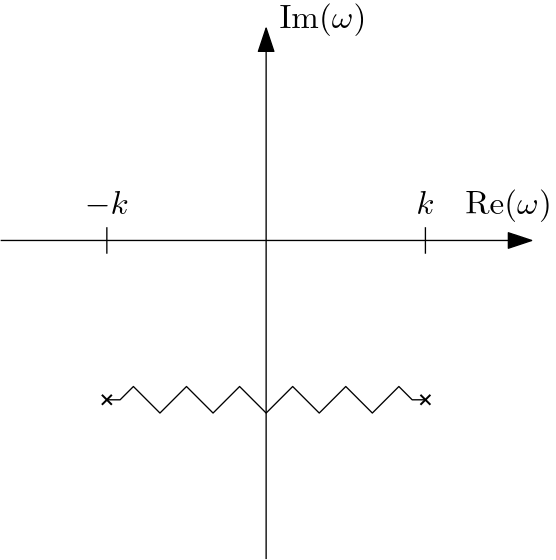}
	\caption{Analytic structure for a function with horizontal branch cut.}
 \vspace{-10pt}
\end{wrapfigure}

We thus find the distribution function has a horizontal logarithmic branch cut with branch points at $\omega=-i/t_R\pm k$. Since this is a one-dimensional non-analyticity, the value of a contour integral containing a section of the branch cut is proportional to the length of this section. The meaningful quantity to look at is therefore this contribution divided by this length, in the limit of this section length going to zero. For a branch cut the resulting value is the discontinuity around a point on the branch cut. If we want to calculate the total signal contribution to the real-time distribution function we have to integrate over this discontinuity, weighted by the factor $e^{-i\omega t}$. In the case of a horizontal branch cut like we have here, the entire branch cut adds a factor decaying at the same rate $e^{-t/t_R}$. On top of that, the fluid can also diffuse in all directions for $k>0$ due to its real part. 

Branch cuts typically occur in any weakly coupled system where particles have a continuum of possibilities to decay to, either spatially or in time \cite{Romatschke_2016, Bajec:2024jez, Moore:2018mma}.

\subsection{Hydrodynamic modes} \label{sec:hydromodes}
So far our system has been purely dissipative. As a result, the entire distribution function decays to zero and no physical quantities are conserved. As an example, we will now look at an adaptation to the collision kernel in (\ref{eq:nhBoltzmann}) that conserves the total number of particles in the system. Thereto the particle density $n(t,\vec{r})=\int f(t,\vec{r},\vec{p}) d\vec{p}$ has to obey the continuity equation $\partial_t n + \vec{\nabla}\cdot(n \vec{v})=0$. Integrating our Boltzmann equation over $\vec{p}$ we find
\begin{equation}
    \frac{\partial n}{\partial t} + \vec{v}\cdot\vec{\nabla}n = \int C[f] d\vec{p} = 0\,.
\end{equation}
This condition can be fulfilled by adding a term $\tilde{f}_{eq}(p)n(\vec{r},t)/t_R$ to the collision kernel, with $\tilde{f}_{eq}$ any thermal distribution function normalized to 1. In Fourier space we now have
\begin{equation}
    f(\omega,\vec{k},\vec{p})= \frac{n(\omega,k)\tilde{f}_{eq}+f(0,k,p) t_R}{1-i\omega t_R+i\vec{v}\cdot\vec{k} t_R}\,.
\end{equation}
Since $n(\omega,k)$ implicitly depends on $f(\omega,\vec{k},\vec{p})$ we need to solve for it first, so we integrate this equation over momentum space to find:
\begin{equation}
    n(\omega,k) = \dfrac{\frac{-i}{2kt_R}\log(\frac{i+\omega t_R-k t_R}{i+\omega t_R+k t_R})}{1+\frac{i}{2kt_R}\log(\frac{i+\omega t_R-k t_R}{i+\omega t_R+k t_R})} n(t=0)t_R\,.
    \label{eq:pdif initial}
\end{equation}
On top of our dissipative branch cut, we now have an additional pole given by the zero of the denominator. Since $k=0$ corresponds to a wave with infinite wavelength and thus a constant shift in the distribution function, these perturbations must be conserved. In terms of the position of the pole, this implies its imaginary part also has to go to zero as $k\rightarrow 0$, resulting in a mode that is arbitrarily long-lived at large wavelengths. For this reason it is called a hydrodynamic mode, while the short-lived dissipative modes are called non-hydrodynamic modes. Generally each conserved quantity will add a hydrodynamic mode, although multiple modes may lie on top of each other due to symmetries in the system. 
These types of modes describe the large scale collective behaviour of fluids at late times and form an essential part of relativistic hydrodynamics \cite{Kadanoff:1963axw, Baier:2007ix}.

\subsection{Non-analytic regions} \label{sec:nas}

\begin{wrapfigure}{R}{0.35\textwidth}
\vspace{-10pt}
	\includegraphics[width=0.35\textwidth,keepaspectratio]{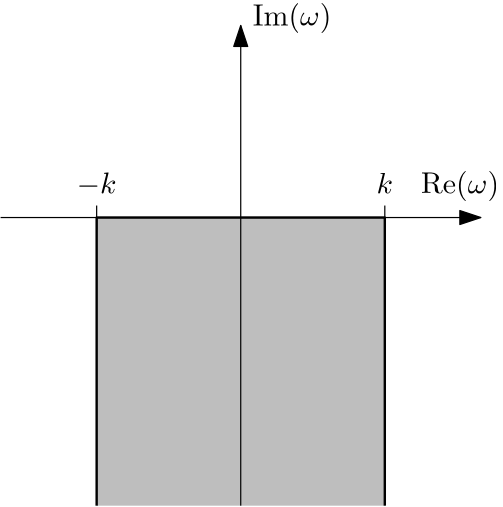}
	\caption{Analytic structure for a function with non-analytic region.}
\vspace{-10pt}
\end{wrapfigure}

So far we have talked about zero-dimensional and one-dimensional analyticities. The next logical step is to look at two-dimensional non-analyticities where a function has non-analytic properties in entire region. So far, to the best of our knowledge, there has been no research on non-analytic regions besides the notion of its existence \cite{Kurkela:2017xis, Ochsenfeld:2023wxz}. Therefore we will have to construct our own definitions and methods along the way. This type of non-analyticity is a lot more subtle in the sense that the function is entirely continuous within this region and all its derivatives can have finite values. In order to find the non-analyticity at a point in this region we have to calculate the value of a contour integral around it. The value of this integral is proportional to the area within it, so when dividing by the surface area $S$ and taking the limit $S\rightarrow 0$ we find a finite value. We therefore define a new property, the {\em non-analytic (area) density}
\begin{equation}
    \nas[f] := \lim_{S\to 0} \left(\frac{1}{S}\oint_{\partial S} f(z)dz\right)
    \label{eq:nas}
\end{equation}
In a non-analytic region the particles necessarily decay at a continuum of decay rates, so we can create such a region by making the relaxation time momentum dependent. This is a very natural assumption, since scattering rates also typically decrease with the momentum of the particle. We therefore propose a relaxation time of the form $\tau_R(p)=t_R p/T$, with $T$ an effective temperature. If we now want to find the particle density for example, we have to integrate (\ref{eq:f branch}) over momentum to find
\begin{equation}
    n(\omega,k) = \frac{-i}{2k}\int_0^\infty p^2 dp f(0,k,p)\log(\frac{iT/p+\omega t_R-k t_R}{iT/p+\omega t_R+k t_R})\,.
\end{equation}

\begin{wrapfigure}{R}{0.3\textwidth}
	\includegraphics[width=0.3\textwidth,keepaspectratio]{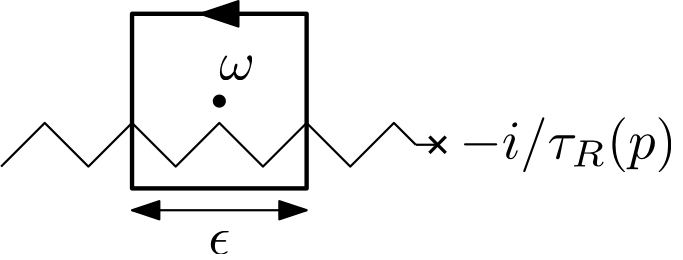}
        \label{fig:nas contour}
\end{wrapfigure}

We now calculate this contour integral for an infinitesimal square with side $\epsilon$ going counter clockwise around $\omega = x-iy$, as shown in the diagram on the right. If we exchange the integrals, we find the contour integral is non-zero only if the branch cut goes through this contour, in which case it is equal to the discontinuity across the branch cut. For a logarithm this is equal to $-\epsilon$Disc$[\log]=-2\pi i\epsilon$. Therefore
\begin{align}
    \vspace{-10pt}
    \nas[n]&=\frac{1}{\epsilon^2}\oint n(\omega,k)d\omega = \frac{-i}{2k\epsilon^2}\int_0^\infty p^2 dp f(0,p,k) \oint  d\omega \log(\frac{iT/p+\omega t_R-k t_R}{iT/p+\omega t_R+k t_R}) \nonumber \\
    &=-\frac{\pi}{k\epsilon}\int_{T/(yt_R+\epsilon t_R)}^{T/(yt_R-\epsilon t_R)} p^2 dp f(0,p,k) = -\frac{\pi}{k} \frac{T^3}{y^4t_R^3}f\left(0,\frac{T}{yt_R},k\right)\,.
    \label{eq:nas n}
\end{align}
This calculation assumes $|\text{Re}(\omega)|<k$, outside of this region the non-analyticity is always zero. As expected, we find a finite value. For distribution functions with an exponential tail in $p/T$, the non-analytic density decreases as $e^{-1/y}$ near the real axis. This still allows for a separation between hydrodynamic modes and the non-hydrodynamic sector, but only at moderately small times. \\
\\
We can conclude that a non-analytic region such as this one is a very general occurrence in physical systems with a continuum of relaxation times. In fact, without any manipulations branch cuts only seem to exist in the limit of $k=0$ or for a system with particles that all thermalize at a single decay rate. This non-analytic region allows the fluid to dissipate in a variety of ways, where the properties of the non-hydrodynamic sector are determined by the distribution of $\nas$ over the complex $\omega$-plane. For example if $f(0,p,k)\propto e^{-p/T}$, equation (\ref{eq:nas n}) has a peak at $y=1/(4t_R)$, indicating an initial exponential decay with effective relaxation time $4 t_R$. \\
\\
This notion of a non-analytic density is entirely new, and so is the method for calculating it in a kinetic theory. Since non-analytic regions are essential for describing the analytic structure of a general kinetic theory, this method already forms the first result of this paper.

\section{Particle diffusion in standard RTA} \label{sec:standard}
Before investigating the general case, let us first go into more detail on the analytic structure of RTA with a constant relaxation time $\tau_R(p)=t_R$. For this discussion the focus will be entirely on particle diffusion where the total number of particles is constant, as the calculations and results are easiest to understand and are qualitatively generalizable to more realistic systems with energy-momentum conservation. The setup we will use is one where the system is constantly perturbed by an electric field $\vec{E}=\vec{\nabla}A_0-\partial_t \vec{A}$ since this allows for a definition of the correlation function that is independent of what other quantities are conserved. The perturbation causes the distribution function to move away from equilibrium by an amount $\delta f=f-f_{eq}$. The Boltzmann equation, linearized in the perturbation, is then given by
\begin{equation} \label{eq:boltzmann standard}
    \left(\frac{\partial}{\partial t}+\vec{v}\cdot\vec{\nabla}\right) \delta f(\vec{r},\vec{p},t) -\frac{\vec{v}\cdot\vec{E}}{T} \feq(p) = \frac{1}{t_R}\left(-\delta f + \frac{\feq}{\chi}\int\frac{d\vec{p}}{(2\pi)^3} \delta f \right) \,,
\end{equation}
where the static susceptibility
\begin{equation}
    \chi=\int\frac{d\vec{p}}{(2\pi)^3}\feq(p)
\end{equation}
normalizes the equilibrium distribution. This additional term of the form $\delta\feq = \delta n (\feq/\chi)$ in the collision kernel is required to guarantee particle conservation as described in section \ref{sec:hydromodes}. Using the same method, we can calculate the particle diffusion correlator
\begin{equation} \label{eq:g00}
    G^{0,0}(\omega,k) \equiv -\frac{\delta n(\omega,k)}{\delta A_0}=-\chi \dfrac{2kt_R+(1-i\omega t_R)iL}{2kt_R+iL}
\end{equation}
with
\begin{equation} \label{eq:L1}
    L\equiv \log(\frac{i+\omega t_R-k t_R}{i+\omega t_R+k t_R})\,,
\end{equation}
a result previously found in \cite{Romatschke_2016}. Note how the correlation function changes compared to (\ref{eq:pdif initial}) where we used an initial perturbation compared to a constant electric field. Although the form of the correlation function changes and therefore the residues change, the locations of the non-analyticities remain identical. 

The same is true when adding energy-momentum conservation under general perturbations. Because the particle density and energy density are coupled, additional hydrodynamic sound modes are added to the correlation function, but the particle diffusion mode can never change its location while remaining in the rest frame. This result is similar to how the correlation function factorizes for calculations at non-zero density \cite{Bajec:2024jez}.

This is an important conclusion as it allows us to study different channels and modes separately without having to worry about our results changing qualitatively when using different perturbations or moving to a system with all conservation laws applied. 

\subsection{Physical picture}

\begin{wrapfigure}{R}{0.4\textwidth}
 \vspace{-10pt}
	\includegraphics[width=0.4\textwidth,keepaspectratio]{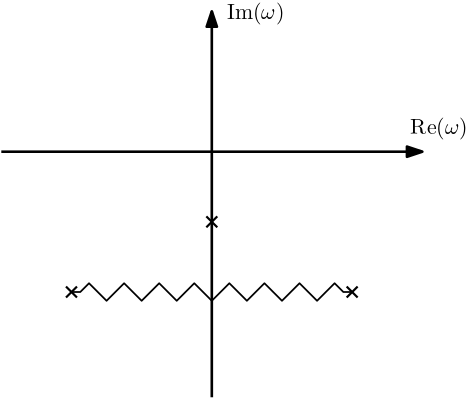}
	\caption{Analytic structure for standard RTA in physical picture.}
	\label{fig:phys stand}
 \vspace{-10pt}
\end{wrapfigure}

Knowing this, we can now start our analysis of the correlation function (\ref{eq:g00}). Due to the ambiguity in the analytic continuation to the complex $\omega$-plane, there are different possibilities as to how the signal is represented, which we will call {\em pictures}. Although the total signal is the same for all of them, how it is distributed over the non-analyticities can differ. Therefore, not all pictures can be interpreted as easily. In this paper, we will focus on two of them that have interesting properties that make them more desirable than others in certain scenarios. One such example of an ambiguity is how the branch points are connected. The most straightforward way of looking at the correlation function is by connecting them with a horizontal branch cut as shown in Figure \ref{fig:phys stand}. This corresponds to integrating over the angle $\theta$ between the $\vec{k}$ vector and the direction $\vec{v}$ of the particle, with $\cos(\theta)$ taking real values between -1 and 1. For this reason we call it the physical picture: it has the most physical meaning behind it and does not require any branch cut manipulations or integration over complex $\cos(\theta)$.

Setting the denominator of (\ref{eq:g00}) to 0 gives a single solution: the location of the diffusion mode
\begin{equation}
    \wh(k) = \frac{-i}{t_R}\left(1-\frac{k t_R}{\tan(k t_R)}\right) = -i\frac{t_R}{3}k^2 + \order{k^4}\,,
\end{equation}
a hydrodynamic mode with diffusivity $D = t_R/3$ \cite{Romatschke_2016}. This solution is not always valid however, because the complex logarithm is a multi-valued function with infinitely many Riemann sheets. When calculating the dispersion relation we have to make a choice of what Riemann sheet we are working on. The requirement of having a hydrodynamic mode determines what our first or principal sheet is. For this $\wh$ to be an actual pole we need the logarithm in (\ref{eq:g00}) to be equal to $2ikt_R$. In the physical picture the imaginary part of the logarithm always lies between $-\pi $ and $\pi $, meaning the pole can only exist when $kt_R$ takes values between $-\pi/2$ and $\pi/2$. Since $1/\tan(kt_R)\rightarrow 0$ at $kt_R=\pi/2$, we can conclude $\wh$ reaches the branch cut at $kt_R=\pi/2$ and moves to the second Riemann sheet, no longer contributing to the real-time signal. 
\\
The residue of this pole is equal to
\begin{equation} \label{eq:resg00 stand}
    \text{Res}(G^{0,0},\wh) = -\chi\wh(k)\frac{k^2t_R^2}{\sin(k^2t_R^2)}\,,
\end{equation}
which is non-zero at $kt_R=\pi/2$, resulting in a discontinuous drop to 0 for the signal of the hydro pole, as a function of $k$. Its time dependence at constant $k$ is a pure exponential decay $\sim e^{-i\wh(k)t}$, but this decay can be very slow at small values of $k$. \\
\begin{figure}[t]
    \begin{subfigure}[]{0.45\textwidth}
            \centering
            \includegraphics[width=\textwidth]{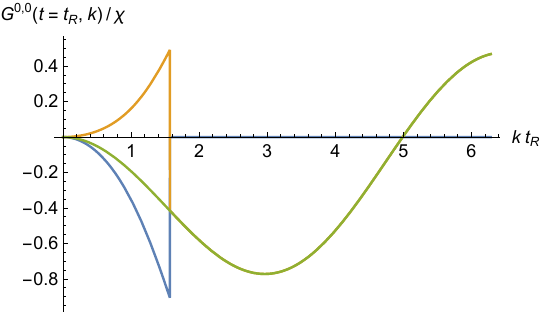}
    \end{subfigure}
    \begin{subfigure}[]{0.55\textwidth}
        \centering
        \includegraphics[width=\textwidth]{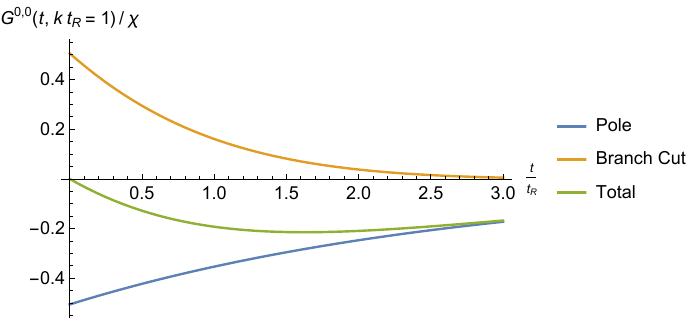}
    \end{subfigure}
    \caption{Signals of different non-analyticities for standard RTA in the physical picture. Left: constant time slice $t=t_R$. Right: constant $k=1/t_R$.}
    \label{fig:phys G00tk}
\end{figure}
\\
The non-analyticity of the branch cut is given by the discontinuity 
\begin{equation}
    \disc[G^{0,0}(-i/t_R+x,k)] = \lim_{\epsilon\to0}\left(G^{0,0}\left(x-\frac{i}{t_R}-i\epsilon,k\right)-G^{0,0}\left(x-\frac{i}{t_R}+i\epsilon,k\right)\right)\,.
\end{equation}
For a logarithm, this discontinuity is $-2\pi i$, so the discontinuity of the denominator is equal to $2\pi$ for $-k<x<k$. Similarly, the discontinuity of the numerator is $-2\pi ix$. Using the quotient rule (\ref{eq:quotient disc}) for discontinuities, we can calculate
\begin{align}
    \disc[G^{0,0}(-i/t_R+x,k)] &= -2\pi\chi \left(\dfrac{-ixt_R}{2kt_R+i\log\left(\frac{k-x}{k+x}\right)+\pi} - \dfrac{2kt_R+xt_R\log\left(\frac{k-x}{k+x}\right)+\pi ixt_R}{\left(2kt_R+i\log\left(\frac{k-x}{k+x}\right)\right)^2-\pi^2} \right) \nonumber \\ 
    &= 2\pi\chi \dfrac{2kt_R(1+ixt_R)}{\left(2kt_R+i\log\left(\frac{k-x}{k+x}\right)\right)^2-\pi^2}\,.
\end{align}
Note that $\disc[G^{0,0}(-i/t_R+x,k)] = \disc[G^{0,0}(-i/t_R-x,k)]^*$, meaning the real part of the discontinuity is symmetric in $x$, while the imaginary part is asymmetric. Its total signal is found by integrating along the entire branch cut:
\begin{equation}
    -\frac{1}{2\pi}\oint_{branch}G^{0,0}(\omega,k)e^{-i\omega t}d\omega = -\frac{1}{2\pi}e^{-t/t_R}\int_{-k}^k \disc[G^{0,0}(-i/t_R+x,k)] e^{-ixt} dx\,,
\end{equation}
which is therefore a real function. As expected, the entire branch cut decays exponentially with relaxation time $t_R$, while the integral adds oscillations corresponding to dampened waves. \\
\\
Finally, we plot the separate signal contributions to $G^{0,0}(t,k)$ in Figure \ref{fig:phys G00tk}. On the left plot, the time is kept constant in order to see what happens when the mode reaches the branch cut at $kt_R=\pi/2$. As the pole disappears, its signal gets absorbed into the branch cut. This causes the branch cut non-analyticity to make a jump too in order for the real-time correlation function to remain constant. 

As a result, for large $k$ only pure dissipation from the branch cut remains. We can interpret this in the following way. At large wavelengths, dissipation is limited by particle number conservation, which introduces the hydrodynamic mode. As the wavelength decreases ($k$ increases), the perturbations can decay faster and $\wh(k)$ moves down in the complex plane. At a critical value of $k t_R=\pi/2$ or wavelength $\lambda = 4 t_R$, the hydro mode reaches the branch cut and the diffusion process is no longer limited by particle number conservation. Therefore, perturbations with smaller wavelengths all decay with relaxation time $t_R$ and a hydro mode is no longer required. This critical value depends on what conservation law is being applied.

In the time dependent plot on the right, the exponential decay with different rates is very clear. Since there is no initial perturbation $G^{0,0}(t=0,k)=0$ the signals have to start opposite. The oscillations of the branch cut are difficult to see in this graph as they only occur for large values of $kt$.

\subsection{Continuous picture}

\begin{wrapfigure}{R}{0.4\textwidth}
\vspace{-10pt}
	\includegraphics[width=0.4\textwidth,keepaspectratio]{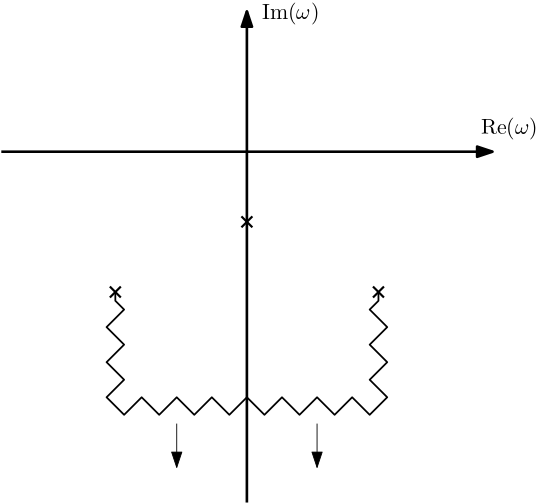}
	\caption{Analytic structure for standard RTA in continuous picture.}
	\label{fig:cont stand}
 \vspace{-10pt}
\end{wrapfigure}

Due to the horizontal branch cut, the diffusion mode disappears at $k t_R = \pi/2$. However, the mode moving to the second Riemann sheet is not necessarily what sets the radius of convergence for hydrodynamic modes, as explained in \cite{Heller:2020hnq} for sound and shear modes. The actual radius of convergence of $\wh(k)$ is set by its single pole at $kt_R=\pi$. In order to reach this divergence we have to deform the branch cut in such a way that it no longer obstructs the mode. This can be done by starting at the branch point at $-k-i/t_R$, moving down to $-k-i\infty$, horizontally to $k-i\infty$, and back up to connect it to the other branch point. This is shown diagrammatically in Figure \ref{fig:cont stand}. Since the horizontal part at $-i\infty$ will not contribute, we end up with 2 vertical branch cuts. 

This choice of branch cuts corresponds to integrating $\cos(\theta)$ from $-1$ to $1$ by passing $-i\infty$ and is therefore less physical. It does however make all non-analyticities continuous as a function of $k$, which is why it is called the continuous picture. To achieve these vertical branch cuts mathematically we have to rotate the argument of each logarithm over a phase $-\pi/2$, resulting in
\begin{equation}\label{eq:L2}
    L = \log(1-i\omega t_R+ik t_R)-\log(1-i\omega t_R-ik t_R)\,.
\end{equation}
The hydro mode still has the same location and residue as it had in the physical picture, but now exists for $kt_R<\pi$. Because we moved the horizontal part of the branch cut to $-i\infty$ the mode only moves to the second Riemann sheet after diverging. The residue (\ref{eq:resg00 stand}) also diverges at $kt_R=\pi$, but due to the exponential suppression $e^{-i\wh(k)t}$ its signal smoothly goes to zero before disappearing. 

\begin{figure}[t]
    \begin{subfigure}[]{0.45\textwidth}
            \centering
            \includegraphics[width=\textwidth]{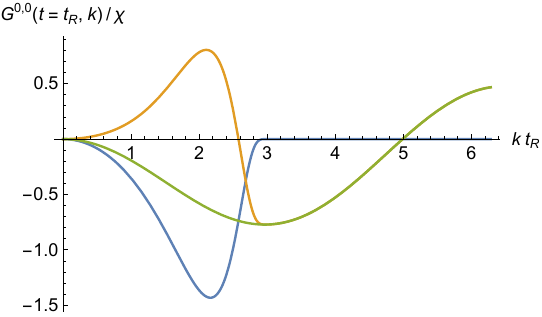}
    \end{subfigure}
    \begin{subfigure}[]{0.55\textwidth}
        \centering
        \includegraphics[width=\textwidth]{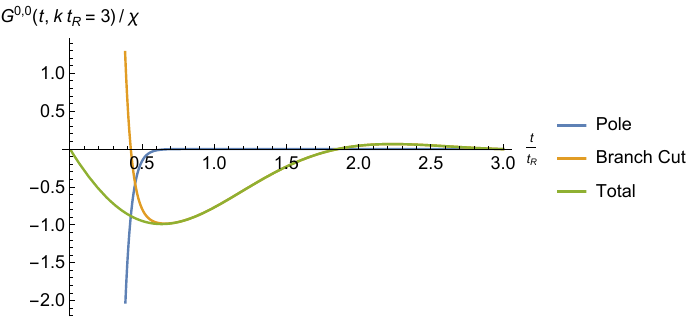}
    \end{subfigure}
    \caption{Signals of different non-analyticities for standard RTA in the continuous picture. Left: constant time slice $t=t_R$. Right: constant $k=3/t_R$.}
    \label{fig:con G00tk}
\end{figure}
The branch cut discontinuities can be calculated in the same way. For the branch at $k$ we find:
\begin{equation} \label{eq:discR}
    \disc[G^{0,0}(k-iy,k)] = -2\pi\chi \dfrac{2k(y+ik)t_R^2}{\left(2kt_R+i\log(yt_R-1)-i\log(1-yt_R-2ikt_R)\right)^2-\pi^2}\,.
\end{equation}
The total signal is given by the sum of the discontinuities along the 2 branch cuts, multiplied by $e^{\pm ikt}$. For convenience we will notate this combined discontinuity as $\Sigma$Disc. For the branch cut at $-k$ we have $\disc[G^{0,0}(-k-iy,k)]=-\disc[G^{0,0}(k-iy,k)]^*$. The combined discontinuity is therefore
\begin{align}
    \Sigma\disc[G^{0,0}] &\equiv e^{-ikt}\disc[G^{0,0}(k-iy,k)]+e^{ikt}\disc[G^{0,0}(-k-iy,k)] \nonumber \\
    &= 2i\,\text{Im}\left(e^{-ikt}\disc[G^{0,0}(k-iy,k)]\right) \,.
\end{align}
For the total signal we now have to integrate in the imaginary direction:
\begin{equation}
    -\frac{1}{2\pi}\oint_{branches}G^{0,0}(\omega,k)e^{-i\omega t}d\omega = \frac{i}{2\pi}\int_{1/t_R}^\infty\Sigma\disc[G^{0,0}]e^{-yt} dy\,.
\end{equation}

The results are plotted in Figure \ref{fig:con G00tk}. For the constant time slice at the left we see there are no more discontinuities. The total correlator is necessarily the same for both pictures at all values of $k$. Since the residue of the pole does not change, the branch cut signal is also the same at $kt_R<\pi/2$. We therefore expect an exponential decay $e^{-t/t_R}$ for these vertical branch cuts. Where this single decay rate arises from is not immediately clear from the position of the branch cut, as we integrate over a continuum of decay rates larger than $1/t_R$. To investigate this further it is useful to look at $kt_R \ll 1$ as the additional oscillations are suppressed in this regime. Expanding (\ref{eq:discR}) we find
\begin{equation}
    \disc[G^{0,0}(k-iy,k)] = \chi\left(yt_R-1+\frac{yt_R^2}{\pi}k + \order{k^2}\right) + i\chi\left(kt_R + \frac{t_R^2}{\pi}k^2 +\order{k^3}\right)\,.
\end{equation}
This $(yt_R-1)$ term seems especially problematic as integrating it over $y$ would result in a $1/t^2$ divergence at small $t$, instead of the pure exponential decay. This issue can be solved through a series of cancellations, the first one being destructive interference between the 2 branch cuts. When summing these, the real part of $\disc[G^{0,0}(k-iy,k)]$ gets suppressed by $\sin(kt)$, resulting in the expansion
\begin{equation}
    \Sigma\disc[G^{0,0}] = 2i\chi\left(1-yt+\frac{t}{t_R}\right)kt_R + \frac{2i\chi}{\pi}(1-yt)k^2t_R^2+\order{k^3}\,.
\end{equation}
This already reduces the divergence to $1/t$. The next type of cancellation results from the integration over $e^{-yt}dy$. Using the gamma function $\Gamma(1)=\Gamma(2)$ we can immediately find that
\begin{equation}
    \int_{1/t_R}^\infty (1-yt)e^{-yt}dy = -\int_{1/t_R}^\infty \frac{t}{t_R}e^{-yt}dy = -\frac{e^{-t/t_R}}{t_R} \,,
\end{equation}
meaning the term at first order in $k$ vanishes and the second order term is simply proportional to $e^{-t/t_R}$. Although we eventually do find the emergence of the exponential decay again, this calculation shows the correct time dependence of this vertical branch cut is a lot harder to resolve, in contrast to the direct interpretation in the physical picture. 

Similarly at $\pi/2 < kt_R < \pi$, the diffusion mode has a large (diverging) residue but also a decay rate $-\text{Im}(\wh(k)) > 1/t_R$. In order to get the same result as the physical picture, $\Sigma\disc[G^{0,0}]$ has a large peak around $y = -\text{Im}(\wh(k))$. This causes the branch cut to rapidly decay at early times too, but since these cancel exactly the sum has a normal $e^{-t/t_R}$ decay. This is shown in the right plot of Figure \ref{fig:con G00tk}. \\
\\
Overall we can conclude the continuous picture with 2 vertical branch cuts is prone to destructive interference and other cancellations. This causes the non-analytic behaviour to be non-intuitive and as a result the interpretation is more complicated compared to the physical picture. It does however make numerical calculations significantly simpler at large $t > t_R$ and $kt_R > \pi$, since the vertical integral is exponentially suppressed while the horizontal integral in the physical picture becomes highly oscillatory. This difference in numerical complexity will become even more significant in momentum dependent RTA.

\subsection{Numerical analytic structure} \label{sec:num ana}
RTA is a useful theory since it allows calculating these non-analyticity in an exact way. However, to investigate the analytic structure of more general and realistic theories, it is practical to have a method that can calculate the value of the non-analyticity numerically. To achieve this we build on the method described in \cite{Ochsenfeld:2023wxz}. They show how a branch cut or non-analytic region can be approximated by a number of poles, and already apply it to standard RTA to calculate the location of the branch cut and hydrodynamic mode. We would like to take this one step further and compare the numerical non-analyticity to the one we can calculate exactly. In the numerical case, the value of a contour integral around a part of the branch cut is given by the sum of the residues of all poles within this contour. In other words, we can use the residues of the poles along the branch cut to approximate the discontinuity around the location of one of these poles. 

\begin{figure}[t]
    \begin{subfigure}[]{0.45\textwidth}
            \centering
            \includegraphics[width=\textwidth]{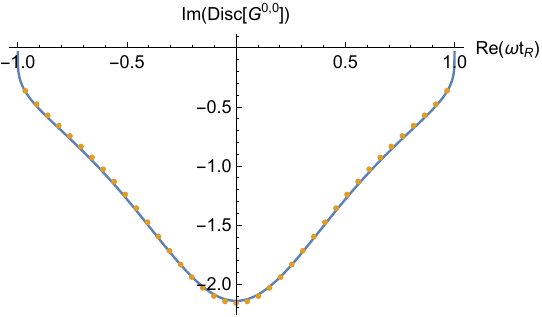}
    \end{subfigure}
    \begin{subfigure}[]{0.55\textwidth}
        \centering
        \includegraphics[width=\textwidth]{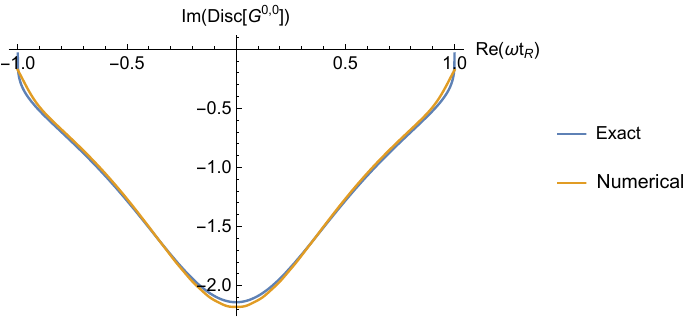}
    \end{subfigure}
    \caption{Comparing numerical calculations of the discontinuity along the horizontal branch cut in standard RTA against the exact results. The momentum grid has $p_{max}=20T$, 100 points in the radial direction and 40 points in the angular direction, with $kt_R=1$. Left: uniform grid with rescaled residues of the poles. Right: calculated by applying a Gaussian kernel to a non-uniform grid.}
    \label{fig:num stand}
\end{figure}

Since this numerical method uses an initial perturbation instead of a constant field, we calculate the discontinuity along the branch cut of (\ref{eq:pdif initial}). As the momentum grid we use is entirely real, the resulting analytic structure coincides with the physical picture, and we end up with a horizontal branch cut:
\begin{equation}
    \disc[G^{0,0}_{\text{init.}}(-i/t_R+x,k)] = \dfrac{4\pi kt_R^2}{\left(2kt_R+i\log\left(\frac{k-x}{k+x}\right)\right)^2-\pi^2}\,.
\end{equation}
For the numerical calculation we need to approximate the momentum integral in (\ref{eq:boltzmann standard}) through a discretization of the  momentum grid. Although the approximation with wedge moments used in \cite{Ochsenfeld:2023wxz} is almost exact in standard RTA, it loses a lot of its accuracy for non-linear integrals in the case of a momentum dependent relaxation time. Instead, we will use an easier midpoint integration method, for example:
\begin{equation}
    \int_0^\infty \frac{p^2}{\tau_R(p)} f(p) dp \equiv \int_0^\infty \frac{N(p)}{\tau_R(p)} dp \approx \sum_{i=0}^{N_p} \frac{N(\Delta p/2 + i\Delta p)}{\tau_R(\Delta p/2 + i\Delta p)} \Delta p \equiv \sum_{i=0}^{N_p} \frac{N_i}{\tau_R(p_i)} \Delta p \,,
\end{equation}
with $N_i$ the $i$-th moment of the distribution function. $\Delta p$ and $N_p$ are the distance between points on, and the size of the momentum grid respectively. The results are qualitatively the same for both integration methods. \\
In order for an infinite series of poles to be equivalent to a branch cut, they need to provide the same signal to the real-time correlation function at all times. This is only possible if the contour integral around any part of the branch cut is equal to the sum of the residues of all poles within this contour, in the equivalent string of poles. To translate the locations $x_i$ and residues $\text{Res}(x_i)$ of the poles into a discontinuity function, we therefore make use of the requirement that a contour integral with length $l$ near $x$ around the series of poles and the branch cut have to yield the same result:
\begin{equation}
    2\pi i\sum_{i=1}^N \text{Res}(x_i) \theta(l/2-|x_i-x|) = \int_{x-l/2}^{x+l/2} \disc[G^{0,0}_{\text{init.}}(-i/t_R+x',k)]dx'\,,
\end{equation}
with $\theta(x)$ the Heaviside step function. If there is a density $\rho(x)$ of poles per unit of length present around $x$ we find that in the limit of $l\to 0$ this equation turns into:
\begin{equation}
    \disc[G^{0,0}_{\text{init.}}(-i/t_R+x_i,k)] = 2\pi i\text{Res}(x_i) \rho(x_i)\,.
\end{equation}
This result can easily be generalized to 2-dimensional non-analyticities where $\rho(x,y)$ is a surface density, as we will show for momentum dependent RTA.

For a uniform distribution of poles $\rho$ is constant and the discontinuity can easily be approximated by interpolating the residues. This is for example the case in RTA when discretizing the angular part of the momentum grid such that it is uniform in $\cos(\theta)$. The analytic and numerical result for this case are compared in the left plot of Figure \ref{fig:num stand}.

Although this works reasonably well, the positions of the poles cannot be fixed directly and small deviations occur, mainly at regions with high non-analyticity. On top of that, even this semi-uniform grid is not always available for more complicated theories. It would therefore be preferable to have a method of calculating the non-analyticity for more general cases. In one dimension $\rho(x)$ can be estimated reasonably easily through the distance of each pole to its closest neighbours. In 2 dimensions however, this is a lot less straightforward. A solution that works in both cases and also allows automatically interpolating the non-analyticity is using a smoothing kernel $K(x,x_i)$ to approximate:
\begin{equation} \label{eq:disc smooth kernel}
    \disc[G^{0,0}_{\text{init.}}(-i/t_R+x,k)] \approx 2\pi i\sum_{i=1}^N \text{Res}(x_i) K(x,x_i)
\end{equation}
A typical example would be the Gaussian kernel
\begin{equation}
    K(x,x_i) = \frac{1}{\sqrt{2\pi}\sigma}\exp\left(-\dfrac{|x-x_i|^2}{2\sigma^2}\right)\,,
\end{equation}
where the prefactor gives the appropriate normalization in one dimension. The standard deviation $\sigma$ determines how many of the nearby poles will significantly contribute to the approximation of the discontinuity. For low values of $\sigma$ the result will oscillate between poles, while for values that are too high some of the behaviour of the non-analyticity may be erased. Sharp edges or oscillations are mostly prone to this. Typically, a value around the maximal distance between the poles gives a good approximation.

To test this result, we discretize our momentum grid such that it is uniform in $\theta$ instead of $\cos(\theta)$. This causes the pole density to heavily increase around the branch points, reducing the residues in this region while increasing them around the middle part. After applying our Gaussian kernel we obtain the plot in the right of Figure \ref{fig:num stand}. \\
\\
Overall, we can conclude this method of numerically calculating the non-analyticities works well, as long as the grid spacing does not vary too much. The numerical values are not fully accurate, but the qualitative behaviour is very clear and sufficient to obtain a physical interpretation of the analytic structure.

\section{Momentum dependent RTA} \label{sec:pdep}
Now that we have a good understanding of how to interpret the analytic structure for a kinetic theory with constant relaxation time, we move on to the general case where particles with different momenta can thermalize over multiple time scales. This results in a relaxation time $\tau_R(p)$ that depends on the momentum of each particle. For a discrete set of relaxation times, this leads to a series of branch cuts. The more interesting case however, is when there is a continuum of relaxation times. This spreads out the branch cut into a non-analytic region that is horizontally bounded by Re$(\omega) = \pm k$ and vertically by the inverse of the minimal and maximal values of $\tau_R(p)$. When the relaxation time has no limits the non-analytic region spans the entire causal region from the real axis to $-i\infty$. 

Because the results are very similar in all channels, we will once again focus on the particle diffusion channel. This channel requires particle number conservation, so the dominant method for thermalization are $2\to2$ scattering processes. In this case the elastic scattering cross section is given~by~\cite{Thomson_2013}
\begin{equation}
    \sigma = \frac{1}{16\pi E_\text{CM}^2}|\mathcal{M}|^2 \,.
\end{equation}
Although the matrix element $\mathcal{M}$ generally has some momentum dependence, most of the momentum dependence of $\sigma$ is captured in the center-of-mass energy $E_\text{CM}$. Since we are interested in the thermalization time scale of a particle with momentum $p$, the particles it will collide with have momentum of the order of the temperature $T$. In this case $E_\text{CM}\sim4pT$ and we find approximately $\sigma \propto 1/p$. As the time between collisions is inversely proportional to the cross section, we expect the relaxation time $\tau_R\propto p$ for thermalization through elastic scattering \cite{Dashthesis,gavassino2024gapless}. To generalize this momentum dependence and investigate the transition from standard RTA, we will look at a power law dependence $\tau_R(p) = (p/T)^\xi t_R$. This relaxation time also encaptures other physical thermalization processes such as radiative energy loss in QCD \cite{Dusling:2009df}. The methods in this paper also work for other functions, as long as $\tau_R(p)$ is invertible and has a (piecewise) continuous derivative. Although the justification of using RTA directly in this scenario can be questioned \cite{Hu:2024tnn}, it provides useful insights into the resulting analytic structure. \\
\\
We also have to adapt the collision kernel slightly to ensure the particle number stays conserved~\cite{Rocha:2021zcw}:
\begin{equation}
    C[\phi] = \frac{p\feq(p)}{\tau_R(p)} \left(-\phi + \dfrac{\expval{p\phi/\tau_R}}{\expval{p/\tau_R}}\right) \,,
\end{equation}
where we use $\delta f = \phi\feq$ and introduce the expectation value
\begin{equation}
    \expval{...} \equiv \int dP \,(...)\, \feq(p) \equiv \int \frac{d^3\vec{p}}{p\,(2\pi)^3} \,(...)\, \feq(p) \,.
\end{equation}
To find the correlation function we start by rearranging the Boltzmann equation to
\begin{equation}
    \phi = \frac{\dfrac{\expval{p\phi/\tau_R}}{\expval{p/\tau_R}}+ \tau_R\vec{v}\cdot\vec{E}}{1-i\omega \tau_R+i(\vec{v}\cdot\vec{k}) \tau_R}\,.
\end{equation}
Since this denominator will appear many times over, it deserves its own notation
\begin{equation}
    \mathcal{D} = 1-i\omega \tau_R+i(\vec{v}\cdot\vec{k}) \tau_R \,.
\end{equation}
Before we can calculate the particle density from this equation, we need to solve for $\expval{p\phi/\tau_R}$. Calculating this expectation value yields
\begin{equation}
    \expval{\frac{p\phi}{\tau_R}} = \frac{\expval{p/(\N\tau_R)}}{\expval{p/\tau_R}}\expval{\frac{p\phi}{\tau_R}} + \expval{\frac{\vec{p}\cdot\vec{E}}{\N}} \,,
\end{equation}
so we find
\begin{equation}
    \phi = \frac{1}{\N}\left(\tau_R\vec{v}\cdot\vec{E} + \dfrac{\expval{\vec{p}\cdot\vec{E}/\N}}{\expval{p/\tau_R}-\expval{p/(\N\tau_R)}}\right) \,.
\end{equation}
The particle density is given by the expectation value $\expval{p\phi}$ and is therefore equal to
\begin{equation}
    n(\omega, k) = \expval{\frac{\tau_R}{\N}\vec{p}\cdot\vec{E}} + \dfrac{\expval{p/\N}\expval{\vec{p}\cdot\vec{E}/\N}}{\expval{p/\tau_R}-\expval{p/(\N\tau_R)}} \,.
\end{equation}
Since our original system is isotropic, we can choose to lay $\vec{k}$ along the $z$-axis. This eventually gives the correlation function
\begin{equation} \label{eq:g00 pdep}
    G^{0,0}(\omega,k) = -\frac{\delta n(\omega,k)}{\delta A_0} = -ik\expval{\frac{\tau_R}{\N}p_z} -ik \dfrac{\expval{p/\N}\expval{p_z/\N}}{\expval{p/\tau_R}-\expval{p/(\N\tau_R)}} \,.
\end{equation}
The expectation values in this correlator can consist of an integral over the angular part of $\vec{p}$ and an integral over $p$. The angular integral can be calculated analytically as we did before for standard RTA. For particle diffusion these relevant integrals are
\begin{align}
    \int\frac{d\Omega}{4\pi}\frac{1}{\N} &= \frac{-i}{2k\tau_R}L\,, \\
    \int\frac{d\Omega}{4\pi}\frac{p_z}{\N} &= \frac{-i}{k\tau_R}\left(1+(1-i\omega\tau_R)\frac{i}{2k\tau_R}L\right),
\end{align}
where $L$ is the same logarithm as previously (\ref{eq:L1}, \ref{eq:L2}), but now has a momentum dependence through $\tau_R$. Due to this logarithm, the integrals over $p$ can not be solved exactly however, and most results will have to be calculated through series expansions or numerical integration.

\subsection{Physical picture}

\begin{wrapfigure}{R}{0.5\textwidth}
\vspace{-20pt}
	\includegraphics[width=0.5\textwidth,keepaspectratio]{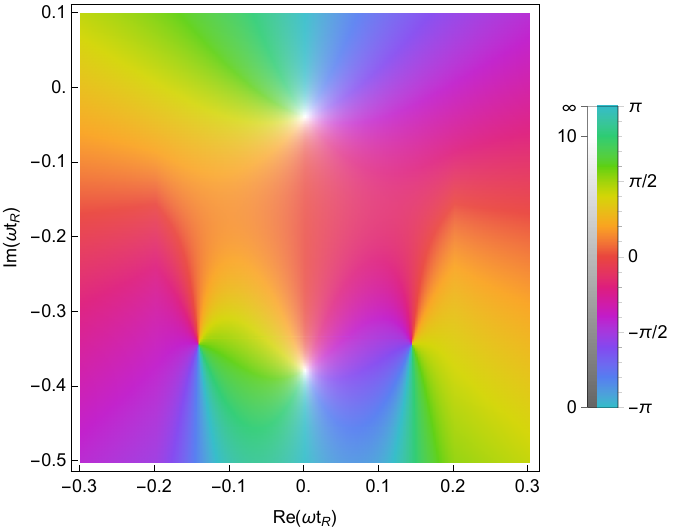}
	\caption{Analytic structure for $\tau_R\propto p$ in the physical picture at $kt_R = 0.2$. The white dots correspond to divergences in the correlation function, while the colors indicate how the argument changes around them.}
	\label{fig:phys pdep}
 \vspace{-20pt}
\end{wrapfigure}

When making the relaxation time momentum dependent, the horizontal branch cut in the physical picture gets smoothened out into a non-analytic region. To see what other structures remain we plot the correlation function (\ref{eq:g00 pdep}) by numerically calculating the integrals, as shown in Figure \ref{fig:phys pdep}. The white dot near the real axis corresponds to the hydrodynamic diffusion mode. The colors correspond to the argument/phase of the correlation function and their behavior around the pole indicates the type of pole. For a single pole around the point $z=re^{i\theta}$ we have $1/z = e^{-i\theta}/r$, so the phase $-\theta$ should rotate around the pole from $-\pi$ to $\pi$ clockwise a single time while the phase of $z$ rotates counter clockwise. The colors on the plot show the hydro mode is indeed a single pole. 
We also see the branch cut in standard RTA gets replaced by a pole and 2 zeroes (dark points) that mimic its behaviour, especially at small values of $\xi$. 

From the colors around the non-hydro mode, we see the phase rotates counter clockwise around it, similar to a pole in the complex conjugate of $z$: $1/z^*=e^{i\theta}/r$. We can thus write that around $\omega=\wnh$:
\begin{equation}
    G^{0,0}(\omega) \approx G^{0,0}(|\omega-\wnh|) e^{i\theta}\,,
\end{equation}
and therefore a circular contour integral with vanishing radius yields
\begin{equation}
    \oint G^{0,0}(\omega) d\omega \approx \int_0^{2\pi} \lim_{\omega\to\wnh}|\omega-\wnh| G^{0,0}(|\omega-\wnh|) e^{2i\theta}id\theta = i\text{Res}(G^{0,0}, \wnh) \int_0^{2\pi}e^{2i\theta}d\theta = 0\,.
\end{equation}
As a result we cannot directly use the residue theorem to calculate its contribution to the contour integral. Using numerical contour integrals around the pole, we find the signal is spread out horizontally and the effective residue is equal to the complex conjugate of Res$(G^{0,0},\wnh)$. The angular dependence of the zeroes also counter-intuitively behaves as that of a pole, but because the residue is zero their contribution remains zero. This odd behaviour of the poles and zeroes can be explained through the non-analyticity of the correlation function in the entire region around them. If the function was analytic, it could be expanded in a Laurent series around $\wnh$ and the lowest order term would be of the form $1/(\omega-\wnh)^n$ with $n>0$ the order of the pole, immediately fixing the rotation of the phase around this pole. Therefore, this kind of behaviour is only possible for non-analytic functions that do not have a Laurent series expansion. 

\subsubsection{Modes}

\begin{figure}[t]
    \centering
    \includegraphics[width=0.7\textwidth]{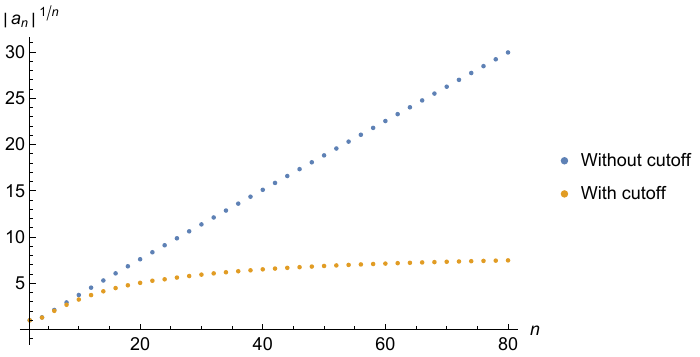}
    \caption{Convergence test by plotting $|a_n|^{1/n}$ as a function of $n$ with and without cutoff on $\tau_R$. The limit of $n\to\infty$ for these values gives the inverse of the convergence radius. In both cases initially $\tau_R\propto p$. The cutoff is applied at $\tau_R=10t_R$.}
    \label{fig:convergences}
\end{figure}

Knowing the non-analyticities we are working with, we start by calculating the location of the modes. These are given by the zeroes of the denominator in (\ref{eq:g00 pdep}):
\begin{equation} \label{eq:mode condition}
    \int  p^2dp\feq(p)\frac{-i}{2k\tau_R(p)^2} \log\left(\frac{i+\omega\tau_R(p)-k\tau_R(p)}{i+\omega\tau_R(p)+k\tau_R(p)} \right) = \int p^2dp \feq(p)\frac{1}{\tau_R(p)}\,.
\end{equation}
In the limit $k\to0$ this equation turns into
\begin{equation}
    \int  p^2dp\feq(p)\frac{1}{\tau_R(p)(1-i\omega(0)\tau_R(p))}  = \int p^2dp \feq(p)\frac{1}{\tau_R(p)}\,.
\end{equation}
As we could already see from the numerical plot, the poles need to have Re$(\omega(0)) = 0$ for the integral to be real. The hydro mode with $\omega(0)=0$ is a trivial solution, and in the case of constant $\tau_R$ it is the only one. The second solution is more subtle, as the integrand has a pole on the real $p$-axis for negative imaginary $\omega(0) = -iy$. By adding a small real part $\omega(0)=-iy\pm\epsilon$ we can use the Sokhotski–Plemelj relation
\begin{equation}
    \lim_{\epsilon\to 0} \frac{1}{x\pm i\epsilon} = \mathcal{P}\left(\frac{1}{x}\right) \mp i\pi\delta(x) \,,
\end{equation}
where $\mathcal{P}$ refers to the Cauchy principal value, to write the integral as ($\tau_R\equiv\tau_R(p)$)
\begin{equation}
    \int  p^2dp\feq\frac{1}{\tau_R(1-y\tau_R\mp i\epsilon\tau_R)}  = \mathcal{P}\int  p^2dp\feq\frac{1}{\tau_R(1-y\tau_R)} \pm i\pi \int  p^2dp\feq\frac{1}{\tau_R}\delta(1-y\tau_R)\,.
\end{equation}
The imaginary part of the integral is simply a result of the branch cut running along the imaginary $\omega$-axis in the case of $k=0$. In this case, the integral has an imaginary part at all $\epsilon\ne 0$ and there is no non-hydro mode. For $k>0$ the vertical branch cut becomes a non-analytic region with a continuous correlation function, meaning the imaginary part vanishes for $\epsilon\to 0$ and the non-hydro mode appears. Its location $\lim_{k\to0}\wnh(k)$ is then given by setting the Cauchy integral equal to the integral on the right-hand side of equation (\ref{eq:mode condition}). This equation generally does not have a solution in closed form, but it can be calculated numerically. 

\begin{figure}[t]
    \begin{subfigure}[]{0.45\textwidth}
            \centering
            \includegraphics[width=\textwidth]{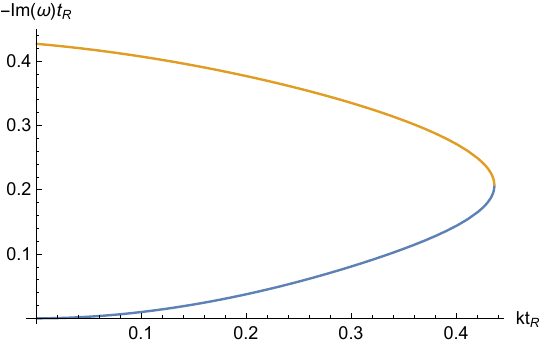}
    \end{subfigure}
    \begin{subfigure}[]{0.55\textwidth}
        \centering
        \includegraphics[width=\textwidth]{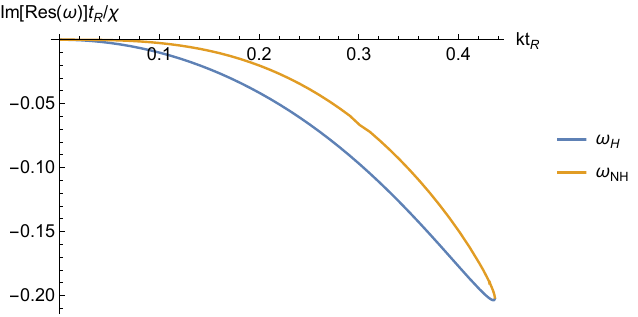}
    \end{subfigure}
    \caption{Dispersion relation (left) and residue (right) of the hydro and non-hydro mode for $\tau_R\propto p$. The modes meet at $kt_R\approx 0.4365$. Note that the residues were calculated as if the poles were analytic. The signal for the non-hydro mode at $t=0$ will be the opposite of this value.}
    \label{fig:modes}
\end{figure}

The hydro mode still looks like an analytic pole and we expect its behaviour to be at least semi-analytic, meaning a Taylor expansion is not exact but does give good results. In the next section we will see this is true because the non-analyticity around the pole is extremely small. We can therefore attempt calculating the behaviour of the hydro mode at small $k$ in the form of a hydrodynamic expansion
\begin{equation}
    \wh(k) = i\sum_{n=1}^\infty a_n k^n \,.
\end{equation}
After filling this into equation (\ref{eq:mode condition}) we can expand the logarithm in a series expansion and solve the equation iteratively for each order in $k$. As is typical for diffusion modes, all odd terms have to be zero. In contrast to standard RTA, the transport coefficients do depend on the choice of equilibrium distribution so we opt for a Boltzmann distribution $\feq(p) = e^{-p/T}$. For $\tau_R(p) = (p/T)^\xi t_R$ we find for the second order transport coefficient the diffusivity $D=-a_2 = \frac{\Gamma(3+\xi)t_R}{6}$. Higher order transport coefficients contain an increasing number of terms. On top of that, the transport coefficients diverge factorially as $|a_n| \sim \Gamma(\beta(\xi)+n\xi)$ with $\beta(\xi)$ a constant in $n$. This makes the hydrodynamic expansion an asymptotic series with a vanishing radius of convergence. This makes sense as the non-analytic region reaches all the way up to the real axis when $\tau_R(p)$ has no maximal value. By setting a cutoff at the relaxation time, for example $\tau_R(p) = \text{Min}(p, 10)$, the non-analytic region becomes gapped as it only reach up to $y=1/10$. Therefore, the hydrodynamic expansion now has a finite convergence radius \cite{Gavassino:2024xwf}. The comparison between these choices of relaxation times is demonstrated in Figure \ref{fig:convergences}. Without the cutoff $|a_n|^{1/n} \propto n$, so from the root test follows the convergence radius is 0. After applying a cutoff this series now converges to a finite value equal to the inverse of the convergence radius. A series expansion for $\wnh(k)$ does not give any meaningful results due to the non-analytic behaviour of the pole.

Finally, in the physical picture for standard RTA the branch cut stopped the hydro mode from going to far into the imaginary plane. In momentum dependent RTA, the non-hydro mode takes care of this. As $k$ increases, they move towards each other and disappear when meeting. The dispersion relations of these modes are calculated numerically in Figure \ref{fig:modes}. We can also calculate the residues of these modes as if the function was analytic:
\begin{equation}
    \text{Res}(G^{0,0}, \omega_*) \equiv \lim_{\omega\to\omega_*}(\omega-\omega_*) G^{0,0}(\omega-\omega_*) \,,
\end{equation}
For an analytic function this corresponds to the coefficient of $1/(\omega-\omega_*)$ in the Laurent expansion. We see in the right plot that these residues become identical when meeting. However, because the effective residue of the non-hydro mode is the complex conjugate of its analytic residue, their signals will be opposite at this point. We can therefore speak of a mode annihilation. 

\subsubsection{Non-analytic region}
Without cutoffs, the non-analytic region spans the entire negative imaginary plane between $\pm k$. Since causality requires all non-analyticities to exist in this region this gives us very little information. What does contain information on the real-time behaviour is how the non-analyticity is distributed within this region. We can describe this using the non-analytic area density $\nas$ defined in (\ref{eq:nas}). For the explicit calculations we will use $\feq = e^{-p/T}$ and $\tau_R(p) = (p/T)^\xi t_R$. 

The correlation function (\ref{eq:g00 pdep}) has 4 parts that have non-analytic properties, that is the expectation values containing $1/\N$. The non-analytic density therefore also has 4 terms. To calculate the separate non-analyticities we use the method described in \ref{sec:nas}. For $\expval{p/\N}$ for example we get
\begin{equation}
    \nas\left[\expval{\frac{p}{\N}}\right] = \nas\left[ \int_0^\infty dp \frac{p^2e^{-p/T}}{2\pi^2} \frac{-iL}{2k(p/T)^\xi t_R} \right] \,,
\end{equation}
where we substitute $u = (p/T)^{-\xi}$:
\begin{align}
    \nas\left[\expval{\frac{p}{\N}}\right] &= \frac{\chi}{2}\nas\left[ \int_0^\infty \frac{du}{\xi u^{3/\xi}}\exp(-u^{-1/\xi}) \frac{-i}{2k t_R}\log\left(\dfrac{i (u-yt_R) + xt_R-k t_R}{i (u-yt_R) + xt_R+kt_R}\right) \right] \nonumber \\
    &= \frac{\chi}{2}\int_0^\infty \frac{du}{\xi u^{3/\xi}}\exp(-u^{-1/\xi}) \frac{-i}{2k t_R}(- 2\pi i t_R) \delta(u-yt_R)\theta(y)\theta(k^2-x^2) \nonumber \\
    &= -\chi\frac{\pi}{2\xi k} \exp(-(yt_R)^{-1/\xi}) \frac{1}{(yt_R)^{3/\xi}}\theta(y)\theta(k^2-x^2) \,.
\end{align}
The first term in the correlation function similarly has non-analytic density
\begin{align}
    \nas\left[\expval{\frac{\tau_R}{\N}p_z}\right] &= \frac{\chi}{2}\int_0^\infty \frac{du}{\xi u^{3/\xi}}\exp(-u^{-1/\xi}) \frac{(1-i\omega t_R/u)}{2k^2 t_R}(- 2\pi i t_R) \delta(u-yt_R)\theta(y)\theta(k^2-x^2) \nonumber \\
    &= -\chi\frac{\pi}{2\xi k} \exp(-(yt_R)^{-1/\xi}) \frac{1}{(yt_R)^{3/\xi}}\frac{x}{yk}\theta(y)\theta(k^2-x^2) \,.
\end{align}
The other 2 non-analytic expectation values can be found by simply multiplying these 2 results with $y$. We can now apply our rules for the non-analytic density of a product/quotient (\ref{eq:product nas},\ref{eq:quotient nas}) to find the total non-analyticity of the correlator
\begin{align}\label{eq:nasg00}
    \nas\left[G^{0,0}(x-iy,k)\right] &= \chi\frac{i\pi}{2\xi} \exp(-(yt_R)^{-1/\xi}) \frac{1}{(yt_R)^{3/\xi}} \biggl\{ \frac{x}{yk} + \dfrac{\expval{p_z/\N}}{\expval{p/\tau_R}-\expval{p/(\N\tau_R)}} \nonumber \\ &+ \frac{x}{k}\dfrac{\expval{p/\N}}{\expval{p/\tau_R}-\expval{p/(\N\tau_R)}} + y \dfrac{\expval{p/\N}\expval{p_z/\N}}{[\expval{p/\tau_R}-\expval{p/(\N\tau_R)}]^2} \biggr\}\theta(y)\theta(k^2-x^2) \,.
\end{align}

\begin{wrapfigure}{R}{0.52\textwidth}
\vspace{-10pt}
	\includegraphics[width=0.5\textwidth,keepaspectratio]{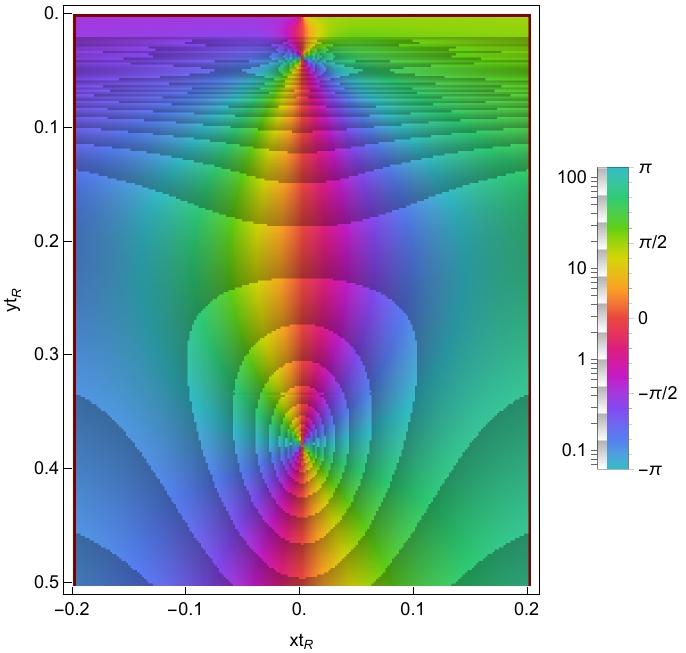}
	\caption{Complex structure of the non-analytic density for $\xi=1$ and $kt_R=0.2$. Since the colors repeat twice around the poles, they are of second order. The series of vertical lines at the top indicate the exponential suppression near the real axis.}
	\label{fig:nas structure}
 \vspace{-10pt}
\end{wrapfigure}

Although the expectation values in the final three terms cause the non-analytic density to have a rather complicated structure, the general behaviour is the same for all terms. At small $y$ near the real axis, the non-analyticity is exponentially suppressed. This is very important since it justifies using a contour integral along the real axis, without having to worry about any non-analyticities. It also explains why the hydrodynamic mode still behaves semi-analytically and has an asymptotic series expansion that gives approximately correct results for a small number of terms. In contrast, the non-hydro mode lives near the peak of the non-analytic density. It therefore behaves completely non-analytically and does not have a correct series expansion. At large $y$ the non-analytic density decreases as a power law $1/y^n$. The value of $n$ determines where the peak in the non-analyticity lies, and therefore the exponential decay rate at moderate times.  

On top of this general behaviour, the final term adds a second order pole at the locations of the hydro mode and non-hydro mode. This can be seen clearly in the complex plot of the non-analytic density in Figure \ref{fig:nas structure}.The non-analyticity around the hydro mode will usually be exponentially suppressed, but this is definitely not the case for the non-hydro mode. Although this may seem problematic at first, when calculating a surface integral over a pole the positive and negative parts around the pole cancel and the total integral takes a finite value. These poles in $\nas$ therefore not necessarily have a major contribution to the surface integral, but they do complicate physical interpretation and numerical calculations. \\
\\
The total signal of the non-analytic region to the real-time correlation function is given by the 2-dimensional integral
\begin{equation}
    -\frac{1}{2\pi}\iint \nas[G^{0,0}(\omega,k)]e^{-i\omega t}dS_\omega = -\frac{1}{2\pi}\int_{-k}^k dx\, e^{-ixt} \int_0^\infty dy\, e^{-yt}\nas[G^{0,0}(x-iy, k)] \,.
\end{equation}
The integral over $x$ yields the oscillatory part of $G(t,k)$, while the vertical integral is responsible for the exponential decay. Therefore, if we want to investigate the exponential decay rate without having to worry about the non-hydro mode it is useful to integrate out the real part of $\omega$ first and plot the result in function of $y$. Since the real part of $G^{0,0}$ is symmetric in $x$ and the imaginary part asymmetric, this integral becomes entirely real:
\begin{equation}
    -\int_{-k}^k \nas[G^{0,0}(x-iy, k)]e^{-ixt} dx = -2\int_0^k \left( \text{Re}[\nas] \cos(xt) + \text{Im}[\nas] \sin(xt) \right) dx \,.
\end{equation}

\begin{wrapfigure}{R}{0.45\textwidth}
\vspace{-10pt}
	\includegraphics[width=0.45\textwidth,keepaspectratio]{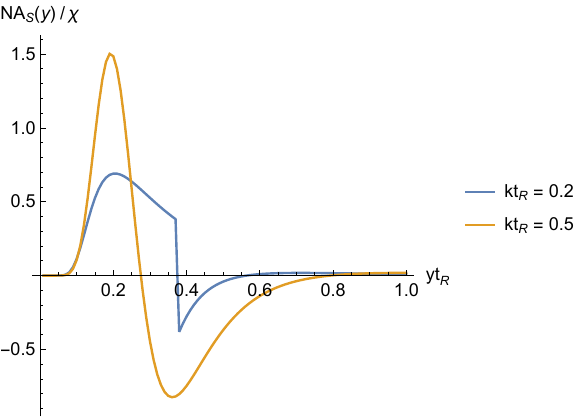}
	\caption{Re$[\nas]$ integrated over $x$, for $\xi=1$. At small $k$, the non-hydro mode causes a discontinuity in the non-analytic density. This jump gets smoothened out after the pole disappears at $kt_R\approx 0.4365$.}
	\label{fig:nasy}
 \vspace{-10pt}
\end{wrapfigure}

At small values of $kt$, the second term is smaller than the first while the first term has very little time dependence. To find the exponential decay at moderately small times it is therefore sufficient to only look at the integral over Re$[\nas]$. The result is plotted in Figure \ref{fig:nasy}. Because some terms in the non-analytic density contain a factor $x/k$, what term is dominant depends on the value of $x$. The location of their peak is also different for each term, so integrating over $x$ results in a wider peak that is a combination of the individual ones.

\begin{figure}[t]
    \begin{subfigure}[]{0.45\textwidth}
            \centering
            \includegraphics[width=\textwidth]{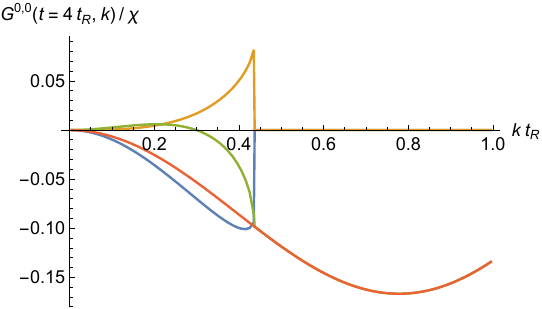}
    \end{subfigure}
    \begin{subfigure}[]{0.55\textwidth}
        \centering
        \includegraphics[width=\textwidth]{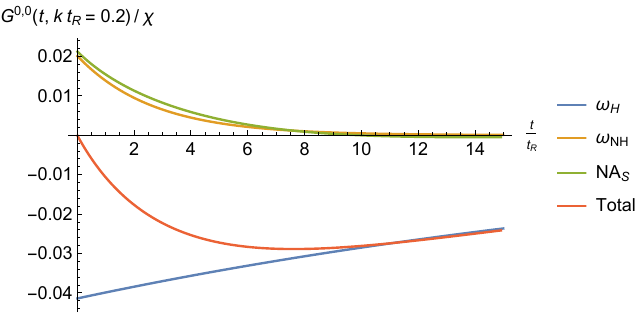}
    \end{subfigure}
    \caption{signals of different non-analyticities for $\tau_R\propto p$ in the physical picture. Left: constant time slice $t=4t_R$, showing the mode annihilation at $kt_R\approx0.4365$. Right: constant $kt_R=0.2$.}
    \label{fig:physp G00tk}
\end{figure}

The non-hydro mode is also important for the non-analytic density as it allows for discontinuities in the terms that have a pole at its location. At early times it is mostly this negative tail that decays, causing the total non-analytic density signal to decrease slower than that of the non-hydro mode. After this initially slow decay where most particles with small momentum thermalize, the non-analytic region starts its exponential decay with a rate determined by the peak in the non-analytic density. The location of the peak is mostly independent from $kt_R$, similar to how the decay rate in standard RTA was fixed. Since the value of $y$ at the peak is smaller than $|$Im$(\wnh)|$, this exponential decay is slower than that of the non-hydro mode. At these moderate time scales the system is governed almost entirely by the hydro mode, indicating the start of the hydrodynamic phase. This process is demonstrated in the left plot of Figure \ref{fig:physp G00tk}. \\
\\
In contrast to standard RTA, the hydrodynamic phase does not last forever \cite{Kurkela:2017xis, gavassino2024gapless}. At any $k>0$, the hydro mode keeps decaying exponentially, so we expect particles with $\tau_R(p) > |1/\text{Im}(\wh)|$ to eventually dominate the correlation function. At these late times, only the part of the non-analytic region with $yt_R \ll 1$ survives under the factor $e^{-yt}$. It is therefore sufficient to just look at the first term in (\ref{eq:nasg00}):
\begin{equation}
    \nas\left[G^{0,0}(x-iy,k)\right] \approx \chi\frac{i\pi}{2\xi} \exp(-(yt_R)^{-1/\xi}) \frac{1}{(yt_R)^{3/\xi}} \frac{x}{yk} \,.
\end{equation}
The integral over $x$ is very straightforward:
\begin{equation}
    \int_{-k}^k \frac{x}{k} e^{-ixt} dx = 2i\frac{\cos(kt)}{t}-2i\frac{\sin(kt)}{kt^2} \,,
\end{equation}
here only the first term matters at large $t$. The integral over $y$ does not have an analytic solution for general $\xi$. However, for rational $\xi$ we can calculate a series expansion at late times. Using this method we propose the following form for the leading order term in the expansion:
\begin{align}
    &\int_0^\infty \exp(-(yt_R)^{-1/\xi}) \frac{1}{(yt_R)^{3/\xi}} \frac{e^{-yt}}{y} dy \nonumber \\
    &=  \exp\left[-\dfrac{\xi+1}{\xi}\left(\frac{\xi t}{t_R}\right)^{\dfrac{1}{\xi+1}}\right] \left[\sqrt{\frac{2\pi}{\xi+1}} \xi \left(\frac{\xi t}{t_R}\right)^{\dfrac{5}{2\xi+2}} + \order{\left(\frac{t}{t_R}\right)^{\dfrac{3}{2\xi+2}}} \right] .
\end{align}
We thus find that for $\xi>0$ the non-analytic region will decay sub-exponentially at late times, and will eventually overtake the hydro mode at $k>0$. However, this can take extremely long for small values of $\xi$ and/or $k$. Altogether, at $t\gg t_R$ we obtain a real-time signal

\begin{equation}
    \frac{\chi}{2} \sqrt{\frac{2\pi}{\xi+1}} \exp\left[-\dfrac{\xi+1}{\xi}\left(\frac{\xi t}{t_R}\right)^{\dfrac{1}{\xi+1}}\right] \left(\frac{\xi t}{t_R}\right)^{\dfrac{5}{2\xi+2}} \dfrac{\cos(kt)}{t} \,.
\end{equation}
\\
Note that the exponential factor is a direct result of the exponential tail in the equilibrium distribution. As a result, this factor does not depend on which correlation function is being investigated and will also be the same for every equilibrium distribution with an exponential tail. 

\newpage
\subsection{Continuous picture}

\begin{wrapfigure}{R}{0.5\textwidth}
 \vspace{-10pt}
	\includegraphics[width=0.5\textwidth,keepaspectratio]{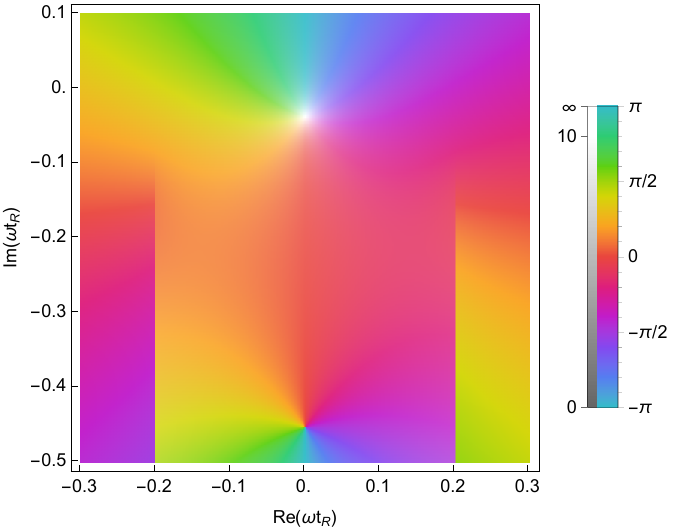}
	\caption{Analytic structure for $\tau_R\propto p$ in the continuous picture at $kt_R = 0.2$. The white dot indicates a pole, the discontinuous change in color a branch cut.}
	\label{fig:con pdep}
 \vspace{-10pt}
\end{wrapfigure}

When transitioning from standard RTA to momentum dependent RTA in the continuous picture, there is no horizontal branch cut to be spread out. Therefore there is also no non-analytic region and the non-hydro sector consists of only 2 vertical branch cuts going all the way up to the real axis as seen in \ref{fig:con pdep}. This makes the continuous picture a lot more efficient at numerically calculating the separate signals from the hydro and non-hydro sector. Such a picture with two vertical branch cuts has previously been researched extensively in \cite{Kurkela:2017xis} for the shear and sound channel, although using a different approach. Therefore this section will mostly focus on what is different in particle diffusion, and compare the two pictures. 

At small $k$, the dispersion relation of the hydrodynamic mode is still the same as it was in the physical picture. In the continuous picture however, there is no non-hydro mode to stop it from reaching $-i\infty$. The value of $k$ when this happens can be calculated exactly using
\begin{equation}
    \lim_{\omega \to -i\infty} L = \lim_{\omega \to -i\infty} \left[ \log(1-i\omega \tau_R+ik \tau_R)-\log(1-i\omega \tau_R-ik \tau_R) \right]= 2\pi i
\end{equation}
for any finite value of $k$ and $\tau_R$. Filling this into our condition for the mode location (\ref{eq:mode condition}) gives
\begin{equation}
    k = \frac{\expval{p/\tau_R^2}}{\expval{p/\tau_R}} \pi = \frac{\Gamma(3-2\xi)}{\Gamma(3-\xi)} \frac{\pi}{t_R}
\end{equation}
for a relaxation time $\tau_R(p) = (p/T)^\xi t_R$. This functional form is rather peculiar since its value is the same at $\xi = 0$ and $\xi=1$ and reaches a minimum in between, at $\xi\approx 0.565$. It then diverges at $\xi = 3/2$ after which the theory is no longer valid. \\
\\
Although calculating the discontinuity along a branch cut is straightforward compared the non-analytic density, finding an analytic solution for the discontinuity is more difficult and in this case not possible for general $\xi$. For example, for the branch cut at $\omega=k-iy$ we find
\begin{equation}
    \text{Disc}\left[\expval{\frac{p}{\N}}\right] = \int_0^\infty dp \frac{p^2e^{-p/T}}{2\pi^2} \frac{-i}{2k(p/T)^\xi t_R} \text{Disc}\left[L\right] = -\chi \frac{\pi}{2 k t_R} \int_0^\infty p^{2-\xi} e^{-p} \theta(y-p^{-\xi}) dp \,.
\end{equation}
For general $\xi$ this results in an incomplete gamma function with no solution in closed form. For $\xi = 1$ we can solve it to be
\begin{equation}
    \text{Disc}\left[\expval{\frac{p}{\N}}\right] = -\chi \frac{\pi}{2 k t_R} \left(1+\frac{1}{yt_R}\right) e^{-1/(yt_R)}
\end{equation}
As expected, the $e^{-1/y}$ factor returns, but we do not find the power law tail we had in the non-analytic density. We will see this is once again a result of the cancellations that occur when having two vertical branch cuts. To find the discontinuity for the entire correlation function we would have to use the product rules for discontinuities. This results in 4 terms that become very complicated since the correlator is not continuous, and are a lot more difficult to interpret compared to the non-analytic density. At small $k$, the leading order terms for the branch cut at $+k$ are
\begin{equation}
    \text{Disc}\left[G^{0,0}(k-iy,k)\right] \approx \chi \frac{\pi}{k}y e^{-1/(yt_R)} + \chi i\pi\left(1+\frac{1}{yt_R}\right)e^{-1/(yt_R)} + ...
\end{equation}
Summing the 2 branch cuts together at small $kt$ gives
\begin{equation}
    \Sigma\disc[G^{0,0}] = \chi 2\pi i e^{-1/(yt_R)} \left(1+\frac{1}{yt_R}-yt\right) + \order{k} \,.
\end{equation}
We can now use the modified Bessel functions $K_n(x)$ to cancel this term:
\begin{equation}
    \int_0^\infty (1-yt) e^{-yt-1/(yt_R)} dy = -\int_0^\infty \frac{1}{yt_R} e^{-yt-1/(yt_R)} dy = -2K_0\left(2\sqrt{t}\right)
\end{equation}
Analogous to standard RTA, we find the vertical branch cuts are prone to cancellations that have to be subtracted first. \\
\\
The general conclusion for the comparison between the pictures is similar to what we found in standard RTA. When making explicit analytic calculations, the physical picture works better in general cases. It also allows for an immediate physical interpretation compared to the continuous picture, as the latter can be deceiving due to its common cancellations. However, when only interested in numerically calculating the separate contributions from the hydro sector and the entire non-hydro sector at small $k$, the continuous picture is more beneficial. 

In the limit of $k=0$ both pictures reduce to a single vertical branch cut. This causes the non-hydro mode to disappear and both pictures become entirely identical. 

\subsection{Shear channel}
In previous parts we have only looked at a system with particle number conservation. However, a physical system typically has conservation of total 4-momentum $p^\mu = (p, \vec{p})$. To achieve this we need to use some different terms in the collision kernel \cite{Rocha:2021zcw}:
\begin{equation}
    C[\phi] = \frac{p\feq(p)}{\tau_R(p)} \left(-\phi + p\dfrac{\expval{p^2\phi/\tau_R}}{\expval{p^2/\tau_R}} + 3\vec{p}\cdot \dfrac{\expval{\vec{p}\,p\phi/\tau_R}}{\expval{p^2/\tau_R}}\right) \,.
\end{equation}
With this kernel we can then calculate the modes in the shear and sound channel using the same method as used for particle diffusion. For the shear channel we get in our notation:
\begin{equation}
    G^{0x,0x} = -ik \expval{\frac{p^2v_x^2v_z\tau_R}{\N}} -3ik \frac{\expval{p^2v_x^2/\N}\expval{p^2v_x^2v_z/\N}}{\expval{p^2/\tau_R}-3\expval{p^2v_x^2/(\N\tau_R)}} \,,
\end{equation}
as previously found \cite{Kurkela:2017xis}. 

\begin{wrapfigure}{R}{0.5\textwidth}
 \vspace{-10pt}
	\includegraphics[width=0.5\textwidth,keepaspectratio]{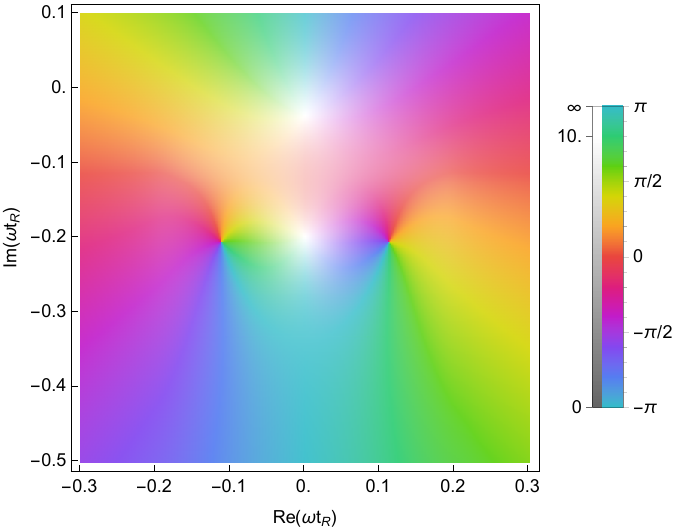}
	\caption{Analytic structure of the shear channel for $\tau_R\propto p$ in the physical picture at $kt_R = 0.2$. The white dots correspond to poles in the correlation function.}
	\label{fig:shear}
 \vspace{-10pt}
\end{wrapfigure}

\begin{figure}[t]
    \centering
    \includegraphics[width=0.7\textwidth]{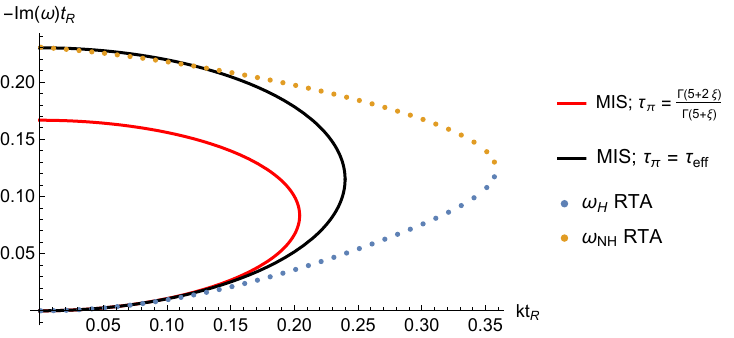}
    \caption{Dispersion relations of the hydro and non-hydro mode in the shear channel. The dots give the numerically calculated values for RTA with $\tau_R \propto p$. The lines give the dispersion relations for MIS with two different values of $\tau_\pi$: the value of the transport coefficient and an effective relaxation time $\teff$.}
    \label{fig:modes shear}
\end{figure}

The analytic structure of this correlation function in the physical picture is shown in Figure \ref{fig:shear}. As expected, the structure is very similar to the one for particle diffusion, but the exact locations of the modes differ. In RTA, the hydro and non-hydro mode move towards each other before annihilating, which stops the modes from becoming acausal. This is similar to Müller-Isreal-Stewart (MIS) theory \cite{muller1967, ISRAEL1976310, ISRAEL1979341}, a UV completion of relativistic hydrodynamics where a non-hydro mode is added, that collides with the hydro mode to keep the theory causal. It is therefore interesting to compare the trajectories of the modes in both theories before they reach each other. 

For the shear channel in MIS theory, the location of the modes is determined by the transport coefficients $\eta$ and $\tau_\pi$. The shear viscosity coefficient $\eta$ we can directly calculate from the hydrodynamic expansion of $\omega_{\text{shear}}(k) = -i\frac{\eta}{sT}k^2+\order{k^4}$, which gives us
\begin{equation}
    \frac{\eta}{s} = \frac{\Gamma(5+\xi)}{120}Tt_R\,.
\end{equation}
On the other hand, $\tau_\pi$ is a bit more subtle. In MIS it plays both the role of a hydrodynamic transport coefficient and of an effective relaxation time $\teff$ of the non-hydrodynamic sector. Its value as a transport coefficient cannot be calculated from the dispersion relation of the shear mode since it mixes with higher order transport coefficients \cite{Grozdanov:2015kqa}, but we can derive it from the dispersion relation of the sound mode to find
\begin{equation}
    \tau_\pi = \frac{\Gamma(5+2\xi)}{\Gamma(5+\xi)}t_R\,.
\end{equation}
Both of these hydrodynamic coefficients also follow directly from the derivation of second-order viscous hydrodynamics from momentum-dependent RTA \cite{Dash:2023ppc}. However, if we plot the MIS dispersion relation with this value of $\tau_\pi$ in Figure \ref{fig:modes shear}, we see the dispersion relation does not match with the non-hydro mode, even for small values of $k$. We can conclude that in RTA, the value of the transport coefficient differs from the effective relaxation time of the non-hydro mode, and one should instead treat $\tau_\pi$ and $\teff$ as separate coefficients. The correspondence between them in MIS is simply a result of the way the theory is constructed. When plotting the dispersion relations using an approximated $\teff$, we see a much better match at small $k$.

\subsection{Sound modes}

\begin{wrapfigure}{R}{0.45\textwidth}
 \vspace{-30pt}
	\includegraphics[width=0.45\textwidth,keepaspectratio]{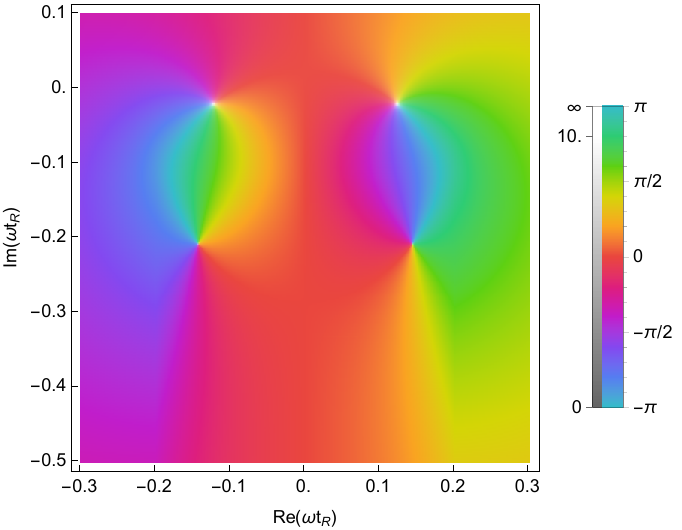}
	\caption{Analytic structure of the sound channel for $\tau_R\propto p$ in the physical picture at $kt_R = 0.2$. The white dots correspond to the locations of the modes.}
	\label{fig:sound}
 \vspace{-20pt}
\end{wrapfigure}

As a result of the conservation of energy and longitudinal momentum, we end up with two more hydrodynamic modes called the sound modes. This is because their real part is non-zero, in contrast to the particle diffusion and shear modes. Instead, they behave as $\omega_{\text{sound}}(k) = \pm c_s k + \order{k^2}$ at small $k$, defining their sound speed $c_s = 1/\sqrt{3}$ for RTA. In momentum-dependent RTA the locations of the modes are given by the solutions of
\begin{equation}
    \expval{\frac{p^2}{\tau_R}\left(1-\frac{1}{\N}\right)} \expval{\frac{p^2}{\tau_R}\left(1-3\frac{v_z^2}{\N}\right)} - 3 \expval{\frac{p^2}{\tau_R}\frac{v_z}{\N}}^2 = 0 \,.
\end{equation}

\begin{figure}[t]
    \centering
    \includegraphics[width=0.7\textwidth]{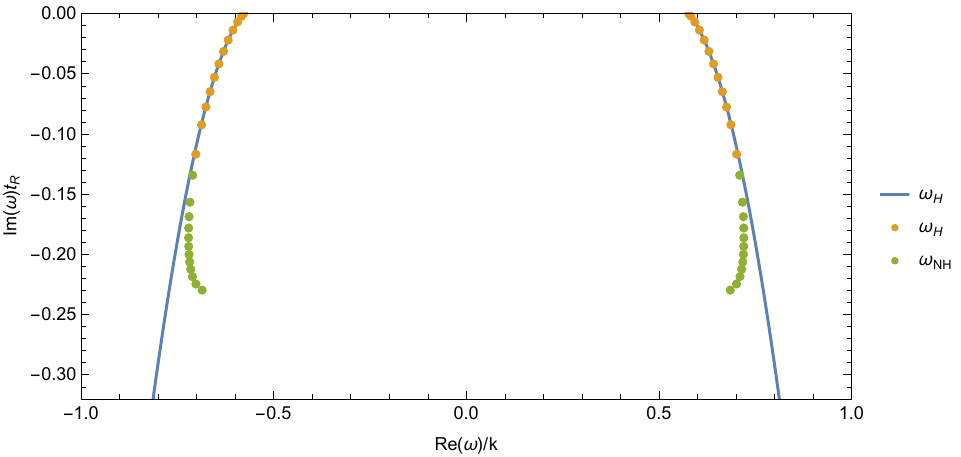}
    \caption{Numerically calculated dispersion relations of the hydro and non-hydro modes in the sound channel for $\tau_R \propto p$. The dots correspond to the physical picture at values of $k$ increasing in steps of $0.5$, until the modes annihilate at $k\approx 0.564$. The lines correspond to the trajectory of the hydro modes in the continuous picture.}
    \label{fig:modes sound}
\end{figure}

These modes are shown in Figure \ref{fig:sound}. There is now a non-hydro mode for each sound mode, who will eventually annihilate with each other. Interestingly, the wave velocity of the non-hydro mode at small $k$
\begin{equation}
    \lim_{k\to 0} \abs{\frac{\wnh(k)}{k}} \approx 0.681
\end{equation}
is larger than the sound speed $c_s \approx 0.577$. In Figure \ref{fig:modes sound} we plot the trajectory of the modes through the complex $\omega$-plane. In the physical picture, the non-hydro mode intercepts the hydro mode at $k\approx 0.564$ before it can reach any large imaginary values. In the continuous picture, there are no non-analyticities for $-k < \Re(\omega) < k$, so both the real and imaginary part of $\wh$ keep increasing in absolute value until the mode reaches the vertical branch cut and gets absorbed there.

\begin{figure}[t]
    \centering
    \begin{subfigure}[]{0.4\textwidth}
            \centering
            \includegraphics[width=\textwidth]{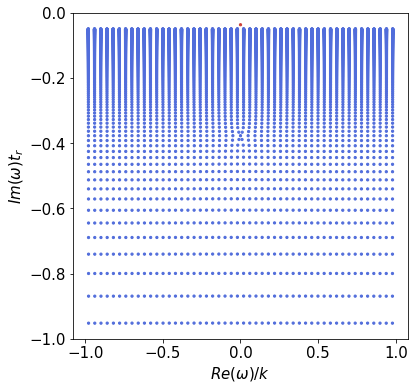}
    \end{subfigure}
    \begin{subfigure}[]{0.43\textwidth}
        \centering
        \includegraphics[width=\textwidth]{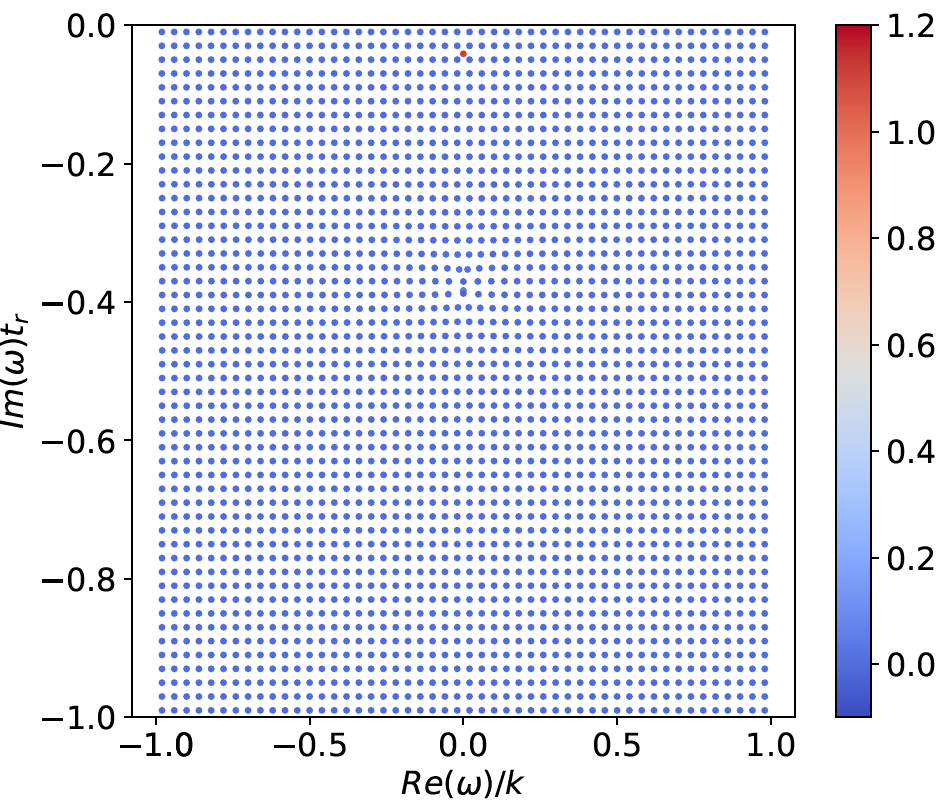}
    \end{subfigure}
    \caption{Numerical calculations of the analytic structure for two types of momentum grids. Both have $\Delta\cos(\theta)=0.04$ and $kt_R=0.2$. Left: grid uniform in $p$, with $\Delta p = 0.1T$ and $p_{max}=20T$. Right: grid uniform in $y=1/\tau_R(p)$, with $\Delta y = 0.02/t_R$ and $y_{max}=8/t_R$.}
    \label{fig:pdep num structure}
\end{figure}

\subsection{Numerical non-analytic region}

In this section, we will generalize the methods from section \ref{sec:num ana} to a 2-dimensional non-analyticity, as it was previously shown such a non-analytic region can be found through numerical means \cite{Ochsenfeld:2023wxz}. Since the numerical calculations are based on a discrete momentum grid, we expect the poles to lie around the locations $\omega_n = \cos{\theta_n}-i/\tau_R(p_n)$. For a uniform grid in $\cos(\theta)$ and $p$, the grid of poles is uniform horizontally but not vertically. The poles lie closer and closer together towards the real axis, and then stop at $y = 1/\tau_R(p_{max})$. This is demonstrated in the left plot of Figure \ref{fig:pdep num structure}. One way of completely avoiding this problem is by writing the equations of motion in the variable $y=1/\tau_R(p)$. This makes the grid of poles indeed almost perfectly uniform as shown on the right of Figure \ref{fig:pdep num structure}, but with the disadvantage that the numerical integrals in $y$ are less accurate for a finite grid compared to those in $p$. This transformation to make the field of poles uniform also does not always exist, and requires some prior knowledge of the analytic structure. 

\begin{figure}[t]
    \begin{subfigure}[]{0.3\textwidth}
            \centering
            \includegraphics[width=\textwidth]{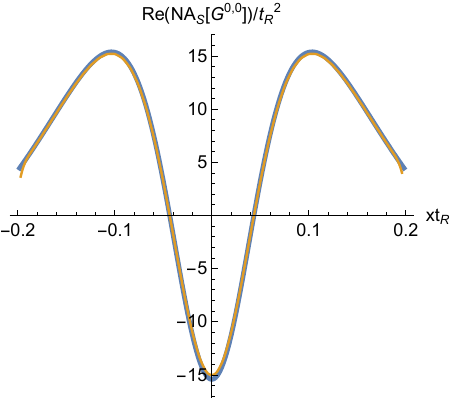}
    \end{subfigure}
    \begin{subfigure}[]{0.3\textwidth}
        \centering
        \includegraphics[width=\textwidth]{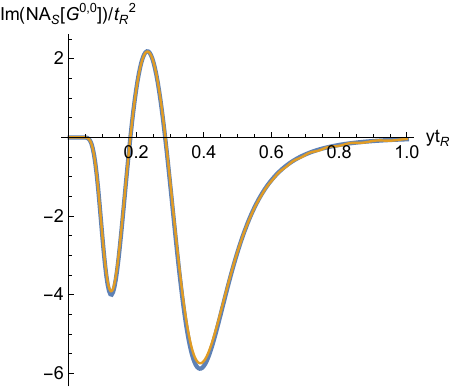}
    \end{subfigure}
    \begin{subfigure}[]{0.4\textwidth}
        \centering
        \includegraphics[width=\textwidth]{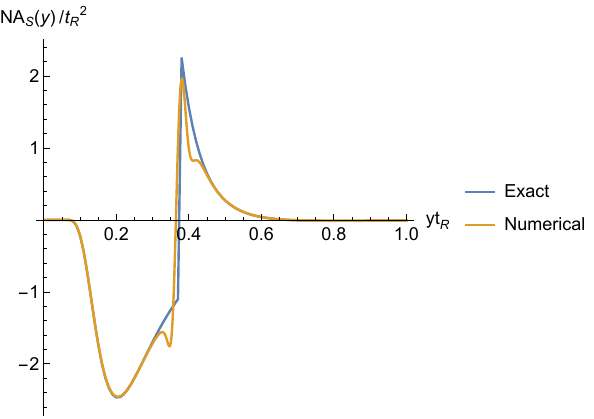}
    \end{subfigure}
    \caption{Comparing numerical calculations of the non-analytic density against the exact results. The momentum grid has $p_{max}=20T$, $\Delta p = 0.1/T$, $\Delta\cos(\theta)=0.04$, and  $kt_R=0.2$. Left: constant $yt_R=0.2$ slice. Middle: constant $xt_R=0.1$ slice. Right: non-analytic density with $x$ integrated out.}
    \label{fig:num nas}
\end{figure}

A better approach is to use a generalization of the kernel method in (\ref{eq:disc smooth kernel}). For a 2-dimensional non-analyticity this is
\begin{equation} \label{eq: nas kernel}
    \nas(x,y) \approx 2\pi i\sum_{i=1}^N \text{Res}(x_i, y_i) K(x,x_i, y, y_i) \,,
\end{equation}
with kernel
\begin{equation}
    K(x,x_i, y, y_i) = \frac{1}{2\pi\sigma_x\sigma_y}\exp\left(-\dfrac{|x-x_i|^2}{2\sigma_x^2}-\dfrac{|y-y_i|^2}{2\sigma_y^2}\right)\,,
\end{equation}
where $\sigma_x$ and $\sigma_y$ can have different values. In fact, for a non-uniform grid such as the one we have here, it is beneficial to vary $\sigma_y$ with $y$. 

To compare the numerical results with an analytic calculation, we have to adapt (\ref{eq:g00 pdep}) to an initial perturbation where $\phi = \phi_0$ is constant. We find
\begin{equation}
    G^{0,0}_{\text{init.}}(\omega,k) = \frac{\delta n(\omega,k)}{\delta n(t=0)} = \frac{1}{\chi} \left[ \expval{\frac{p\tau_R}{\N}} + \dfrac{\expval{p/\N}^2}{\expval{p/\tau_R}-\expval{p/(\N\tau_R)}} \right] \,,
\end{equation}
with non-analytic density
\begin{align}
    \nas&\left[G^{0,0}_{\text{init.}}(x-iy,k)\right] = -\frac{\pi}{2\xi k} \exp(-(yt_R)^{-1/\xi}) \frac{1}{(yt_R)^{3/\xi}} \nonumber \\ &\times \biggl\{ \frac{1}{y} + 2\dfrac{\expval{p/\N}}{\expval{p/\tau_R}-\expval{p/(\N\tau_R)}} + y \dfrac{\expval{p/\N}^2}{[\expval{p/\tau_R}-\expval{p/(\N\tau_R)}]^2} \biggr\}\theta(y)\theta(k^2-x^2) \,.
\end{align}
These results are compared for a horizontal and vertical slice in Figure \ref{fig:num nas}. When not too close to the location of the modes, and in a region with a sufficient amount of poles, the results agree very well. When moving towards large values of $y$, the poles become sparse and small errors start occurring. Close to the non-hydro mode, this method also does not work well. This is because the signal of the non-hydro mode is not a single pole, but is spread out over a small number of poles around it. It is therefore not possible to simply remove the non-hydro pole like we can for the hydro mode. When the location and residue of this pole are known exactly, we can correct for it by subtracting its contribution directly from the sum in (\ref{eq: nas kernel}). Although this improves the approximation near the pole tremendously, there are still some errors due to the interactions between the non-hydro mode and the non-analytic area around it. This can be seen in the right plot of Figure \ref{fig:num nas}, where we integrate out the $x$ dependence of the non-analytic density.

\section{Conclusions and outlook}
Signals contributing to real-time correlation functions can generally originate from three types of non-analyticities: poles, branch cuts and non-analytic regions. Of these three, the non-analytic region is the most general while the others can be seen as limiting cases. In this paper, we found kinetic theory can be used to describe all three of them. For RTA it is also possible to calculate what their non-analytic densities are. Although it is in theory always possible to deduce their signal from the non-analyticity, we have also demonstrated this can be misleading. 

That brings us to the different pictures, a result of the ambiguity in connecting the branch points. We identified two of them that have interesting properties: the physical picture which requires the momentum of the particles to remain real when integrating over it, and the continuous picture that needs the correlation function to be analytic for $-k < \Re(\omega) < k$, not including hydro modes, such that there are no discontinuities in the signal. For RTA, we found the physical picture to have a more direct interpretation. This is likely because the momentum vector is in this case entirely real. In standard RTA, the discontinuity along the horizontal branch cut has the same sign everywhere, so there are no cancellations besides the interference from integrating over $e^{-ixt}$, that only cause oscillations on top of the exponential decay. For momentum dependent RTA the non-hydro mode slightly complicates the interpretation of the non-analytic density, but in practice has little effect on it. The interplay between the hydro and non-hydro mode is very clear, something that is completely missing from other pictures.

On the other hand, the continuous picture is prone to both interference between the vertical branch cuts and also to cancellations due to ambiguity in the vertical integration, for all types of RTA. It is however much easier to calculate the signals numerically, especially in the case of momentum dependent RTA. We can conclude that when only interested in the difference between the hydrodynamic and non-hydrodynamic signals the continuous picture is more appropriate, while the physical picture is required to calculate the signals from the non-analytic region and non-hydro mode separately. 

Since the requirements the two pictures need to satisfy are entirely independent of the theory, we expect the previous conclusions based on them to be general and applicable as long as these pictures can be constructed. \\
\\
We also found a method for analytically calculating the non-analytic density. This is essential to the analytic study of any general theory. The generalizability of this specific method to other theories as of now is not clear. The method is based on the assumption that the non-analytic region can be written as a continuum of branch cuts. In theory, this should be possible by writing the correlation function as an integral over a varying branch cut. In practice however, it is important to have a good understanding of the analytical properties of the integrand. This is for example the case when the integral over either the angular part of $\vec{p}$ or over $|\vec{p}|$ has a solution in closed form. It would be interesting to see if a method exists that can be used to analytically calculate the non-analytic density for a wide range of correlation functions. \\
\\
Another approach to estimate the analytic density is the numerical method based on \cite{Ochsenfeld:2023wxz}. By approximating the analytic structure through a field of poles, we can effectively calculate the non-analytic density for any time-independent linearized Boltzmann equation. This method works well as long as there are no singularities in the density nearby. However, the non-hydrodynamic mode interacts heavily with the non-analytic region, making it difficult to locate the mode and resolve the non-analytic density around it. To improve the results, it can be beneficial to already have a basic understanding of the analytic structure of the theory but this is not strictly necessary. Because of the low requirements of this method, it can also be applied to more physical kinetic theories, such as an ab initio approach with a collision kernel based on effective field theory \cite{Berges:2020fwq, Arnold:2002zm}. \\
\\
Furthermore, we have only discussed linear response in this paper. The next step would be to investigate how the non-analyticities behave and interact with each other in the case of nonlinear hydrodynamics. A good starting point would for example be RTA for Bjorken flow \cite{Heller:2018qvh}. In \cite{Du:2022bel} it was found that an effective relaxation time emerges from QCD effective kinetic theory. Therefore, we expect momentum dependent RTA with Bjorken flow to yield recognizable results. Although the methods in this paper can not directly be applied to this nonlinear theory, our results still form a basis for such an analysis. \\
\\
When the non-analytic region reaches all the way up to the real $\omega$ axis, it has a sub exponential decay at very late times. As a result it will eventually become the dominant signal compared to a hydro mode at $k>0$. In infinitely large systems where $k$ can become arbitrarily small however, this is not very relevant as the resulting hydrodynamic signal to the correlator $G(\vec{r}, t)$ will have a power law decay. In systems with small sizes, we do expect the non-hydro sector to be much more relevant at late times. An example of such small systems are the collisions of protons with other protons or light nuclei. For these systems there are also other factors such as thermal fluctuations that play an important role \cite{Noronha:2024dtq}, so it would be interesting to see how these factors interact.

\section*{Acknowledgements}
This paper was written under the supervision of Michal P. Heller, who I want to thank for all the support, ideas and feedback that improved the result tremendously. I also want to thank the other researchers from the UGent hep-th group for their feedback and the valuable discussions we had, namely Alexandre Serantes, Fabio Ori, Tim Schuhmann and Clemens Werthmann. \\
\\
This project has received funding from the European Research Council (ERC) under the European Union’s Horizon 2020 research and innovation programme (grant number: 101089093 / project acronym: High-TheQ). Views and opinions expressed are however those of the authors only and do not necessarily reflect those of the European Union or the European Research Council. Neither the European Union nor the granting authority can be held responsible for them.

\appendix
\numberwithin{equation}{section}
\section{Product rules for non-analyticities} \label{ap: rules}
In Section \ref{sec:nas} we explained how to calculate the non-analytic density in momentum dependent RTA for a correlation function containing a single integral. However, the full correlation function for a kinetic theory with conserved currents such as \ref{eq:g00 pdep} contains multiple non-analytic integrals that are added and multiplied together. We therefore have to derive a few rules to calculate the full non-analyticity based on that of the individual terms. \\
\\
We start with the discontinuity, defined as
\begin{equation}
    \text{Disc}[f] := \lim_{\epsilon\to 0^+}\left[f(z+\epsilon)-f(z-\epsilon)\right] \,,
\end{equation}
although it can also be interpreted as a non-analytic linear density
\begin{equation}
    \text{Disc}[f] \equiv \lim_{\ell\to 0} \left( \frac{1}{\ell} \oint_\ell f(z)dz \right) \,,
\end{equation}
where the contour integral contains a section of the branch cut with length $\ell$. These interpretations are equivalent if $f(z)$ is analytic around the branch cut. For non-analytic functions, one first has to make the contour integral infinitely narrow around the branch cut, before this interpretation can be applied. From the first definition we can immediately see that
\begin{align}
    \text{Disc}[f+g] &= \lim_{\epsilon\to 0^+}\left[f(z+\epsilon) + g(z+\epsilon) - f(z-\epsilon) - g(z-\epsilon)\right] \nonumber \\
    &= \text{Disc}[f]+\text{Disc}[g] \,.
\end{align}
As a result, any continuous terms can immediately be ignored. For products of 2 functions the rule is less trivial:
\begin{align}
    \text{Disc}[f\cdot g] &= \lim_{\epsilon\to 0^+}\left[f(z+\epsilon)g(z+\epsilon) - f(z-\epsilon)g(z-\epsilon)\right] \nonumber \\
    &= \lim_{\epsilon\to 0^+}\left[f(z+\epsilon)g(z+\epsilon) - f(z-\epsilon)g(z+\epsilon) + f(z-\epsilon)g(z+\epsilon) - f(z-\epsilon)g(z-\epsilon)\right] \nonumber \\
    &= \lim_{\epsilon\to 0^+}\left[\text{Disc}[f]\cdot g(z+\epsilon) + f(z-\epsilon)\cdot\text{Disc}[g]\right] \nonumber \\
    &= \text{Disc}[f]\cdot [g_+] + [f_-]\cdot\text{Disc}[g] \,,
\end{align}
where we introduced a new notation on the last line for future convenience. Since $f$ and $g$ are commutative in this rule, it automatically follows that also
\begin{equation}  \label{eq:product disc}
    \text{Disc}[f\cdot g] = \text{Disc}[f]\cdot [g_-] + [f_+]\cdot\text{Disc}[g] \,,
\end{equation}
or by adding both product rules together and dividing by 2 we find the symmetric product rule:
\begin{equation}
    \text{Disc}[f\cdot g] = \text{Disc}[f]\cdot \frac{[g_-]+
    [g_+]}{2} + \frac{[f_-]+[f_+]}{2} \cdot\text{Disc}[g] \,.
\end{equation}
This rule is especially convenient when the real/imaginary part of $f$ or $g$ is anti-symmetric around the branch cut, since it will drop from the equation. From these product rules we can derive the quotient rule:
\begin{align}
    0 = \text{Disc}\left[\frac{g}{g}\right] &= \frac{\text{Disc}[g]}{[g_+]} + [g_-]\cdot \text{Disc}\left[\frac{1}{g}\right] \nonumber \\
    \Rightarrow \text{Disc}\left[\frac{1}{g}\right] &= -\frac{\text{Disc}[g]}{[g_-]\cdot[g_+]} \nonumber \\
    \Rightarrow \text{Disc}\left[\frac{f}{g}\right] &= \frac{\text{Disc}[f]}{[g_+]} -\frac{[f_+]\cdot\text{Disc}[g]}{[g_-]\cdot[g_+]} \nonumber \\
    &= \frac{\text{Disc}[f]\cdot[g_-] - [f_+]\cdot\text{Disc}[g]}{[g_-]\cdot[g_+]}  \label{eq:quotient disc}\,.
\end{align}
Once again, the subscript $-$ and $+$ can be swapped or symmetrized. \\
\\
Deriving the product rule for the non-analytic area density
\begin{equation}
    \nas[f\cdot g] := \lim_{S\to 0} \left(\frac{1}{S}\oint_{\partial S} f(z)g(z)dz\right)
\end{equation}
is in a sense easier because the functions $f$ and $g$ are continuous in the point $z_0$ where we want to calculate the non-analytic density. We parameterize the contour integral over $z$ around this point as $z(\alpha) = z_0+r(\alpha)\tilde{z}(\alpha)$, where $r = |z-z_0|$ is infinitesimally small, as we will later take the limit $S\to 0$. Using the continuity, we write $f(z(\alpha)) = f(z_0) + r(\alpha)\tilde{f}(\alpha) + \order{r^2}$ and analogous for $g$. Due to the non-analyticity of $f$ we cannot say much about $\tilde{f}$, besides it having no direct dependence on $r$. 

The contour integral then becomes
\begin{equation}
    \oint f(z)g(z)dz = \int_{\alpha_1}^{\alpha_2} \tilde{z}'(\alpha)d\alpha \left( f(z_0)g(z_0) r + \tilde{f}(\alpha)g(z_0) r^2 + f(z_0)\tilde{g}(\alpha) r^2 + \order{r^3} \right)\,.
\end{equation}
When dividing by $S\sim r^2$ and taking the limit $S\to 0$ all higher order terms will vanish. We thus end up with
\begin{align} \label{eq:product nas}
    \nas[f\cdot g] &= \lim_{S\to 0} \frac{1}{S} \left(f(z_0)g(z_0)\oint_{\partial S}dz + g(z_0)\oint_{\partial S}f(z)dz + f(z_0)\oint_{\partial S}g(z)dz\right) \nonumber \\
    &= \nas[f]\cdot g + f \cdot \nas[g] \,.
\end{align}
This is very similar to the product rule for discontinuities, but because $f$ and $g$ are generally continuous functions, there is no need to take right and left limits. Analogous to the derivation for discontinuities, the quotient rule for non-analytic densities is 
\begin{equation} \label{eq:quotient nas}
    \nas\left[\frac{f}{g}\right] = \frac{\nas[f]\cdot g - f \cdot \nas[g]}{g^2} \,.
\end{equation}

\printbibliography

\end{document}